\documentclass[letterpaper,11pt]{article}

\usepackage{cite}
\usepackage[all]{xy}
\usepackage[dvips]{graphicx}
\usepackage{graphpap}
\usepackage{ifthen}
\usepackage{enumerate}
\usepackage{amssymb}
\usepackage{mathrsfs}
\usepackage[errorshow]{tracefnt}
\usepackage{fancybox}
\usepackage{amsfonts}
\usepackage{amsmath}
\usepackage{amssymb}
\usepackage{multirow}
\usepackage{array}
\usepackage{mathtools}
\usepackage{soul}
\usepackage{pict2e}
\usepackage{mathabx}
\DeclareMathAlphabet{\mathpzc}{OT1}{pzc}{m}{it}
\newcommand{\pzc}{\mathpzc}

\makeatletter
\newcommand{\thickhline}{%
    \noalign {\ifnum 0=`}\fi \hrule height 1pt
    \futurelet \reserved@a \@xhline
}
\newcolumntype{'}{@{\hskip\tabcolsep\vrule width 1pt\hskip\tabcolsep}}
\makeatother
\newcolumntype{"}{@{\hskip\tabcolsep\vrule width 1.5pt\hskip\tabcolsep}}
\makeatother

\newcommand{\scr}{\mathscr}

\def\boxit#1{\vbox{\hrule\hbox{\vrule\kern3pt
             \vbox{\kern3pt#1\kern3pt}\kern3pt\vrule}\hrule}}

\newcommand{\bor}{\scr B}

\newcommand{\beq}{\begin{equation}}
\newcommand{\beqn}{\begin{equation*}}
\newcommand{\eeq}{\end{equation}}
\newcommand{\eeqn}{\end{equation*}}
\newcommand{\beqa}{\begin{eqnarray}}
\newcommand{\beqan}{\begin{eqnarray*}}
\newcommand{\eeqa}{\end{eqnarray}}
\newcommand{\eeqan}{\end{eqnarray*}}
\newcommand{\bdm}{\begin{displaymath}}
\newcommand{\edm}{\end{displaymath}}

\newcommand{\la}{\langle}
\newcommand{\ra}{\rangle}

\newcommand{\ba}{\begin{array}}
\newcommand{\ea}{\end{array}}

\newcommand\nn{\nonumber}

\newcommand\benu{\begin{enumerate}}
\newcommand\eenu{\end{enumerate}}
\newcommand\bit{\begin{itemize}}
\newcommand\eit{\end{itemize}}

\def\der'{\mathfrak{der}'\,}
\def\der{\mathfrak{der}\,}
\def\str'{\mathfrak{str}'\,}
\def\str{\mathfrak{str}\,}

\def\frake{\mathfrak{e}}

\def\so{\mathfrak{so}}

\def\sl{\mathfrak{sl}}
\def\gl{\mathfrak{gl}}

\newcommand{\al}{\alpha}
\newcommand{\be}{\beta}

\newcommand{\de}{\delta}

\newcommand{\dlb}{\ensuremath{[\![}}
\newcommand{\drb}{\ensuremath{]\!]}}

\RequirePackage{hyperref}

\numberwithin{equation}{section}

\begin{document}

\frenchspacing

\vskip-10pt
\hfill {\tt \today} \\
\vskip-10pt
\hfill {\tt MI-TH-1520} \\
\vskip-10pt
\hfill {\tt 1507.08828} \\

\pagestyle{empty}

\vspace*{3.5cm}

\begin{center}
\noindent
{\LARGE {\sf \textbf{Exceptional geometry and
Borcherds superalgebras}}}\\
\vspace{.3cm}

\renewcommand{\thefootnote}{\fnsymbol{footnote}}

\vskip 1truecm

\noindent
{\large {\sf \textbf{Jakob Palmkvist}
}}\\
\vskip .5truecm
        {\it 
        {
        Mitchell Institute for Fundamental Physics and Astronomy\\
        Texas A\&M University\\ College Station, TX 77843, USA}\\[3mm]}
        {\tt jakobpalmkvist@tamu.edu} \\
\end{center}

\vskip 1cm

\centerline{\sf \textbf{
Abstract}}
\vskip .2cm

\noindent
We study generalized diffeomorphisms in exceptional geometry with U-duality group $E_{n(n)}$ 
from an algebraic point of view. By extending the Lie algebra
$\frake_n$ to an infinite-dimensional Borcherds superalgebra,
involving also the extension to $\frake_{n+1}$, 
the generalized Lie derivatives can be expressed in a simple way, and the expressions take the same form for any $n\leq7$.
The closure of the transformations then follows from the Jacobi identity and 
the grading of $\frake_{n+1}$ with respect to $\frake_{n}$.

\newpage

\pagestyle{plain}

\section{Introduction}\label{Introduction}

Exceptional geometry generalizes ordinary geometry in eleven-dimensional supergravity, or M-theory, compactified to $D=(11-n)$ dimensions.
In this generalization the $n$-dimensional internal tangent space,
considered as a vector module of ${\rm GL}(n)$, is
extended to an irreducible module
of the U-duality group $E_{n(n)}$. 
All internal bosonic degrees of freedom are unified into a generalized metric,
and the ordinary diffeomorphisms are
unified with tensor gauge transformations
into generalized diffeomorphisms \cite{Hull:2007zu,Pacheco:2008ps,
Hillmann:2009ci,Berman:2010is,Berman:2011pe,Berman:2011cg,
Berman:2011jh,Coimbra:2011ky,Coimbra:2012af,Park:2014una,Park:2013gaj,
Cederwall:2013naa,Cederwall:2013oaa,
Aldazabal:2013mya,
Aldazabal:2013via,Hohm:2013vpa,Hohm:2013uia,Hohm:2014fxa,Hohm:2014qga,Rosabal:2014rga,Cederwall:2015ica}.

The idea presented in this paper is to consider the $E_{n(n)}$ module
as an odd subspace of a Borcherds superalgebra $\bor_{n}$, which is an infinite-dimensional extension of the Lie algebra $\frake_n$,
related to the ordinary Lie algebra extension $\frake_{n+1}$.
By further extending $\bor_{n}$ and $\frake_{n+1}$ to the Borcherds superalgebra $\bor_{n+1}$
we find simple algebraic expressions for
the generalized 
diffeomorphisms which take the same form for any $n\leq7$.
The closure of the transformations
then follows from the Jacobi identity in $\scr B_{n+1}$ 
and the $\mathbb{Z}$-grading of $\frake_{n+1}$ with respect to $\frake_n$.

It is well known already
that the level decomposition of the Borcherds superalgebra $\scr B_n$ with respect to the $\frake_n$ subalgebra gives the correct
spectrum of $p$-forms in maximal $D$-dimensional supergravity, including 
all duals of lower rank fields, and also all additional $(D-1)$- and $D$-form potentials allowed by supersymmetry
\cite{Cremmer:1998px,HenryLabordere:2002dk,HenryLabordere:2002xh,Henneaux:2010ys,Greitz:2011da,Greitz:2012vp}. 
All their equations of motion and Bianchi identities can, using $\scr B_n$, be combined into one Maurer-Cartan equation
and one twisted self-duality relation  
\cite{Cremmer:1998px}, and this result can furthermore be generalized to gauged supergravity
\cite{Greitz:2013pua,Howe:2015hpa},
modifying $\scr B_n$ to a tensor hierarchy algebra \cite{Palmkvist:2013vya}.
The spectrum can also be derived from the indefinite Kac-Moody algebra $\frake_{11}$
\cite{Riccioni:2007au,Bergshoeff:2007qi,Bergshoeff:2007vb,Riccioni:2007ni}
in accordance with the description of M-theory as a nonlinear realization of $\frake_{11}$ \cite{West:2001as}.
The correspondence between
$\scr B_{n}$ and $\frake_{11}$
has been studied in \cite{Henneaux:2010ys,Palmkvist:2011vz},
and generalized to other Borcherds superalgebras and Kac-Moody algebras in \cite{Palmkvist:2012nc,Howe:2015hpa}.

The appearance of the Borcherds superalgebras $\scr B_n$ in the context of exceptional geometry was observed in
\cite{Berman:2012vc}, where it was shown that the generalized diffeomorphisms are reducible, and lead to an infinite tower of ghosts for ghosts.
The corresponding infinite sequence of $\frake_n$-representations agrees precisely with the level decomposition of 
$\scr B_n$ for positive levels, which was later explained in \cite{Cederwall:2015oua}.
The same representations also appear in the tensor hierarchies considered in \cite{Hohm:2013vpa,Hohm:2013uia,Hohm:2014fxa},
related to those appearing in gauged supergravity \cite{deWit:2005hv,deWit:2008ta}.

In this paper we show that the Borcherds structure is in fact hidden already in the
generalized diffeomorphisms themselves,
not only in their reducibility,
and the correspondence with the ghost structure can then be shown directly.
Our results may lead to a way of including gravitational degrees of freedom in the Borcherds approach to
supergravity, as well as to deeper insights into exceptional geometry.
Since our results are generic for
$n \leq 7$ they may
provide some guidance in dealing with the difficulties associated to the dual graviton in the case $n=8$
\cite{Hohm:2014fxa,Rosabal:2014rga,Cederwall:2015ica},
and in
proceeding to $n\geq9$. Unless otherwise stated, we assume
$3 \leq n \leq 7$ in this paper, but
we will also comment on the case $n=8$.

The paper is organized as follows. In section 2 we describe the algebras that we will use, and how they are related to each other.
In section 3 we first review the construction of the generalized diffeomorphisms given in \cite{Berman:2012vc} and
then present our new algebraic expressions.
To derive the identities needed for closure of the transformations
we need to introduce the concept of generalized Jordan triple systems,
and define a superversion of it.
In the end of section 3 we tentatively discuss the reducibility of the transformations in view of our new results,
and we continue the discussion in section~4, where we also point out some natural directions for further research. 

\section{The algebras}

In this section we will define the 
Borcherds superalgebra $\bor_n$ as an extension of the Lie algebra $\frake_n$,
and describe how both $\bor_n$ and $\frake_{n+1}$ can be further extended to $\bor_{n+1}$.
Borcherds (super)algebras are generalizations of Kac-Moody (super)algebras and can themselves be further
generalized to contragredient Lie (super)algebras.
We will only consider the special cases that we are interested in here,
and refer to \cite{Kac77B,Kac,Wakimoto,Ray} for details and general definitions.

We will assume the base field to be the real numbers, so that we get 
the split real forms of the corresponding complex algebras, since these are the ones that appear in the
physical applications that we are interested in. From a purely mathematical point of view,
we can equally well
let the algebras remain complex.

We recall that $\frake_n$, as a special case of a Kac-Moody algebra, is defined as the Lie algebra generated by $3n$ elements $e_i$, $f_i$ and
$h_i=[e_i,f_i]$ ($i=1,2,\ldots,n$) modulo the Chevalley-Serre relations 
\begin{align} \label{chev-rel}
[h_i,e_j]&=a_{ij}e_j, & [h_i,f_j]&=-a_{ij}f_j, & [e_i,f_j]&=\delta_{ij}h_j,
\end{align}
\begin{align} \label{serre-rel}
(\text{ad }e_i)^{1-a_{ij}}(e_j)&=(\text{ad }f_i)^{1-a_{ij}}(f_j)=0 \qquad\quad(i\neq j),
\end{align}
where the Cartan matrix $a_{ij}$ is given by the following Dynkin diagram.
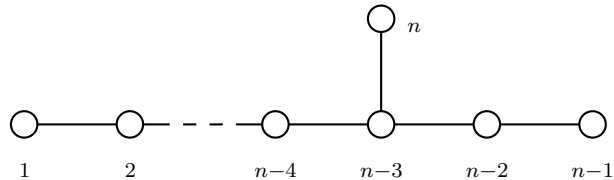
\begin{center}
\scalebox{1}{
\begin{picture}(225,65)(110,-10)
\put(113,-10){${\scriptstyle{1}}$}
\put(153,-10){${\scriptstyle{2}}$}
\put(202,-10){${\scriptstyle{n-4}}$}
\put(242,-10){${\scriptstyle{n-3}}$}
\put(282,-10){${\scriptstyle{n-2}}$}
\put(322,-10){${\scriptstyle{n-1}}$}
\put(260,45){${\scriptstyle{n}}$}
\thicklines
\multiput(210,10)(40,0){4}{\circle{10}}
\multiput(215,10)(40,0){3}{\line(1,0){30}}
\put(155,10){\circle{10}}
\put(115,10){\circle{10}}
\put(120,10){\line(1,0){30}}
\multiput(160,10)(35,0){2}{\line(1,0){10}}
\multiput(175,10)(10,0){2}{\line(1,0){5}}
\put(250,50){\circle{10}} \put(250,15){\line(0,1){30}}
\end{picture}}\end{center}
The nodes represent simple roots $\alpha_i$ which we normalize by 
$(\alpha_i,\alpha_i)=2$.
Their mutual inner products are then either $(\alpha_i,\alpha_j)=-1$ or $(\alpha_i,\alpha_j)=0$, in such a way that
the nodes $i$ and $j$ are connected by
$|(\alpha_i,\alpha_j)|$ lines,
and the Cartan matrix is then given by
\begin{align} \label{cart-mat}
a_{ij}=(\al_i,\al_j).
\end{align}

In this paper we only consider the cases $3\leq n \leq 8$ (and if not $n=8$ is stated explicitly we assume $3\leq n \leq 7$).
For $n=5$ and $n=4$, the Lie algebras $\frake_n$ are not exceptional, but coincide with the classical Lie algebras
$\mathfrak{d}_5=\so(5,5)$ and $\mathfrak{a}_4=\mathfrak{sl}(5)$, respectively.
For $n=3$ the Lie algebra
is not even simple 
but equal to the direct sum of $\mathfrak{a}_1=\mathfrak{sl}(2)$ and 
$\mathfrak{a}_2=\mathfrak{sl}(3)$. 

\subsection{From $\frake_n$ to $\scr B_n$ and $\frake_{n+1}$, and further to $\scr B_{n+1}$}

By adding a node to the Dynkin diagram, $\frake_n$ can be extended to either $\frake_{n+1}$ or to the Borcherds superalgebra $\bor_n$, depending on whether
the node is white (like the orginal nodes in the $\frake_n$ diagram) or gray ($\otimes$) as illustrated below, and as we will now explain.
The additional node represents an additional simple root which we denote by $\alpha_0$
in the $\frake_{n+1}$ case, and by $\beta_0$
in the $\bor_{n}$ case. They have the same
inner product with the simple root $\alpha_1$ of the $\frake_n$ subalgebra,
\begin{align}
(\alpha_0,\alpha_1)=(\beta_0,\alpha_1)=-1
\end{align}
and are orthogonal to all other simple roots of $\frake_n$. However,
like the original simple roots of the $\frake_n$ subalgebra,
$\alpha_0$ has
norm squared equal to two, $(\alpha_0,\alpha_0)=2$, whereas $\beta_0$ is a null root,
$(\beta_0,\beta_0)=0$. Furthermore, $\alpha_0$ is even whereas $\beta_0$ is odd, which means
that the Chevalley generators $e_0$ and $f_0$ associated to $\beta_0$ are odd elements in the Lie superalgebra $\bor_n$ (so that,
for example, $[e_0,f_0]=[f_0,e_0]$
instead of $[e_0,f_0]=-[f_0,e_0]$).
Both $\frake_{n+1}$ and $\bor_{n}$ are then defined
by the same Chevalley-Serre relations as before, (\ref{chev-rel})--(\ref{serre-rel}),
with the Cartan matrix still given by (\ref{cart-mat}),
but now including also the additional simple root $\al_0$ or $\be_0$.
\noindent
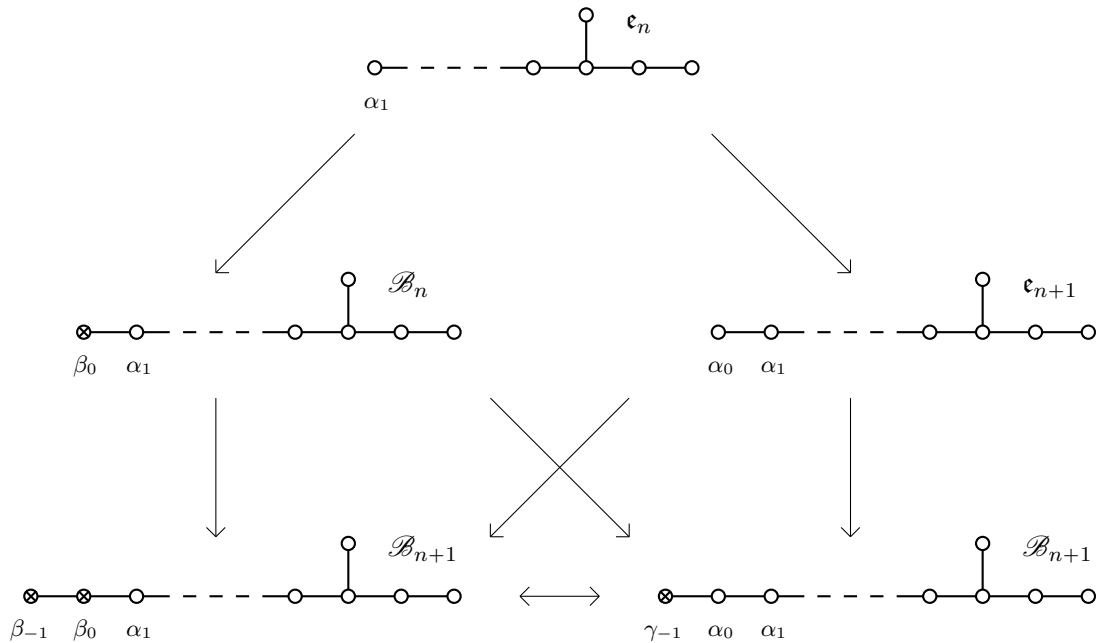
\begin{center}
\begin{picture} (405,240)(17.5,-15)
\thicklines
\put(18,-2){\line(1,1){4}}
\put(22,-2){\line(-1,1){4}}
\put(38,-2){\line(1,1){4}}
\put(42,-2){\line(-1,1){4}}
\put(20,0){\circle{5}}
\put(40,0){\circle{5}}
\put(60,0){\circle{5}}
\put(120,0){\circle{5}}
\put(140,0){\circle{5}}
\put(160,0){\circle{5}}
\put(180,0){\circle{5}}
\put(140,20){\circle{5}}
\put(22.5,0){\line(1,0){15}}
\put(42.5,0){\line(1,0){15}}
\put(62.5,0){\line(1,0){10}}
\put(77.5,0){\line(1,0){5}}
\put(87.5,0){\line(1,0){5}}
\put(97.5,0){\line(1,0){5}}
\put(107.5,0){\line(1,0){10}}
\put(122.5,0){\line(1,0){15}}
\put(142.5,0){\line(1,0){15}}
\put(162.5,0){\line(1,0){15}}
\put(140,2.5){\line(0,1){15}}
\put(258,-2){\line(1,1){4}}
\put(262,-2){\line(-1,1){4}}
\put(260,0){\circle{5}}
\put(280,0){\circle{5}}
\put(300,0){\circle{5}}
\put(360,0){\circle{5}}
\put(380,0){\circle{5}}
\put(400,0){\circle{5}}
\put(420,0){\circle{5}}
\put(380,20){\circle{5}}
\put(262.5,0){\line(1,0){15}}
\put(282.5,0){\line(1,0){15}}
\put(302.5,0){\line(1,0){10}}
\put(317.5,0){\line(1,0){5}}
\put(327.5,0){\line(1,0){5}}
\put(337.5,0){\line(1,0){5}}
\put(347.5,0){\line(1,0){10}}
\put(362.5,0){\line(1,0){15}}
\put(382.5,0){\line(1,0){15}}
\put(402.5,0){\line(1,0){15}}
\put(380,2.5){\line(0,1){15}}
\put(12,-15){\footnotesize
{$\beta_{-1}$}}
\put(36,-15){\footnotesize
{$\beta_0$}}
\put(56,-15){\footnotesize
{$\alpha_1$}}
\put(252,-15){\footnotesize
{$\gamma_{-1}$}}
\put(276,-15){\footnotesize
{$\alpha_0$}}
\put(296,-15){\footnotesize
{$\alpha_1$}}
\put(155,15){{$\bor_{n+1}$}}
\put(395,15){{$\bor_{n+1}$}}
\put(38,98){\line(1,1){4}}
\put(42,98){\line(-1,1){4}}
\put(40,100){\circle{5}}
\put(60,100){\circle{5}}
\put(120,100){\circle{5}}
\put(140,100){\circle{5}}
\put(160,100){\circle{5}}
\put(180,100){\circle{5}}
\put(140,120){\circle{5}}
\put(42.5,100){\line(1,0){15}}
\put(62.5,100){\line(1,0){10}}
\put(77.5,100){\line(1,0){5}}
\put(87.5,100){\line(1,0){5}}
\put(97.5,100){\line(1,0){5}}
\put(107.5,100){\line(1,0){10}}
\put(122.5,100){\line(1,0){15}}
\put(142.5,100){\line(1,0){15}}
\put(162.5,100){\line(1,0){15}}
\put(140,102.5){\line(0,1){15}}
\put(155,115){{$\bor_{n}$}}
\put(280,100){\circle{5}}
\put(300,100){\circle{5}}
\put(360,100){\circle{5}}
\put(380,100){\circle{5}}
\put(400,100){\circle{5}}
\put(420,100){\circle{5}}
\put(380,120){\circle{5}}
\put(282.5,100){\line(1,0){15}}
\put(302.5,100){\line(1,0){10}}
\put(317.5,100){\line(1,0){5}}
\put(327.5,100){\line(1,0){5}}
\put(337.5,100){\line(1,0){5}}
\put(347.5,100){\line(1,0){10}}
\put(362.5,100){\line(1,0){15}}
\put(382.5,100){\line(1,0){15}}
\put(402.5,100){\line(1,0){15}}
\put(380,102.5){\line(0,1){15}}
\put(36,85){\footnotesize
{$\beta_0$}}
\put(56,85){\footnotesize
{$\alpha_1$}}
\put(276,85){\footnotesize
{$\alpha_0$}}
\put(296,85){\footnotesize
{$\alpha_1$}}
\put(395,115){{$\frake_{n+1}$}}
\put(150,200){\circle{5}}
\put(210,200){\circle{5}}
\put(230,200){\circle{5}}
\put(250,200){\circle{5}}
\put(270,200){\circle{5}}
\put(230,220){\circle{5}}
\put(152.5,200){\line(1,0){10}}
\put(167.5,200){\line(1,0){5}}
\put(177.5,200){\line(1,0){5}}
\put(187.5,200){\line(1,0){5}}
\put(197.5,200){\line(1,0){10}}
\put(212.5,200){\line(1,0){15}}
\put(232.5,200){\line(1,0){15}}
\put(252.5,200){\line(1,0){15}}
\put(230,202.5){\line(0,1){15}}
\put(146,185){\footnotesize
{$\alpha_1$}}
\put(245,215){{$\frake_n$}}
\thinlines
\put(142.5,175){\line(-1,-1){52.5}}
\put(90,122.5){\line(1,0){5}}
\put(90,122.5){\line(0,1){5}}
\put(277.5,175){\line(1,-1){52.5}}
\put(330,122.5){\line(-1,0){5}}
\put(330,122.5){\line(0,1){5}}
\put(90,22.5){\line(-1,1){3.54}}
\put(90,22.5){\line(1,1){3.54}}
\put(330,22.5){\line(-1,1){3.54}}
\put(330,22.5){\line(1,1){3.54}}
\put(90,22.5){\line(0,1){52.5}}
\put(330,22.5){\line(0,1){52.5}}
\put(205,0){\line(1,0){30}}
\put(205,0){\line(1,1){3.54}}
\put(205,0){\line(1,-1){3.54}}
\put(235,0){\line(-1,1){3.54}}
\put(235,0){\line(-1,-1){3.54}}
\put(193.75,22.5){\line(0,1){5}}
\put(193.75,22.5){\line(1,0){5}}
\put(193.75,22.5){\line(1,1){52.5}}
\put(246.25,22.5){\line(0,1){5}}
\put(246.25,22.5){\line(-1,0){5}}
\put(246.25,22.5){\line(-1,1){52.5}}
\end{picture}
\end{center}

Both $\frake_{n+1}$ and $\bor_n$ can in turn be extended to the Borcherds superalgebra $\scr B_{n+1}$, as illustrated above.
When we extend $\scr B_n$ to $\scr B_{n+1}$ (left vertical arrow)
we add another odd null root $\beta_{-1}$ to the set of simple roots, such that $(\beta_{-1},\beta_{0})=1$
(note the sign!) and $\beta_{-1}$ is orthogonal
to all the simple roots of $\frake_n$. The embedding of $\frake_{n+1}$ in $\scr B_{n+1}$ is then given by identifying
$\alpha_0$ in $\frake_{n+1}$ with $(\beta_{-1}+\beta_0)$ in $\scr B_{n+1}$ (diagonal arrow going to the left).
Alternatively, we could extend $\frake_{n+1}$ to $\scr B_{n+1}$ by adding an odd null root $\gamma_{-1}$ such that $(\gamma_{-1},\alpha_{0})=-1$
and $\gamma_{-1}$ is orthogonal to the simple roots of $\frake_n$. The embedding of $\bor_{n}$ in $\scr B_{n+1}$ would then
be given by identifying
$\beta_0$ in $\bor_{n}$ with $(\gamma_{-1}+\alpha_0)$ in $\scr B_{n+1}$, as described in \cite{Kleinschmidt:2013em}.
The two different Dynkin diagrams of $\scr B_{n+1}$ are related by a so-called odd reflection
mapping $\gamma_{-1}$ and $-\beta_{-1}$ to each other \cite{Kleinschmidt:2013em,Howe:2015hpa}.
We choose the former approach here, corresponding to the left Dynkin diagram of $\scr B_{n+1}$,
so that the subscripts $0$ and $-1$ of any Chevalley generators always refer to
$\beta_0$ and $\beta_{-1}$.

The Killing form on $\frake_n$ can be extended to a supersymmetric invariant bilinear form on the whole of
$\bor_{n+1}$, which we denote by $\la x |y\ra$ for any two elements 
$x$ and $y$. Supersymmetry here means $\la x | y \ra =-\la y|x \ra$ if both elements
are odd, and
$\la x | y \ra =\la y|x \ra$ if at least one of them 
is even. Invariance always means
$\la [x,y] | z \ra = \la x | [y,z] \ra$.

\subsection{Level decompositions}

The extension of $\frake_n$ by the additional simple root $\beta_0$ gives rise to a $\mathbb{Z}$-grading, or {\it level decomposition}, of $\bor_n$ 
with $e_0$ and $f_0$ at level $+1$ and $-1$, respectively, and the
$\frake_n$ subalgebra at level zero (together with the Cartan element 
$h_0=[e_0,f_0]$).
For any integer $p$ we denote the subspace of $\bor_n$ at level $p$ by $\scr U_p$, so that $[\scr U_p,\scr U_q] = \scr U_{p+q}$, and $\bor_n$
is the direct sum of all these subspaces.
In the same way, the extension of $\frake_n$ by the additional simple root $\alpha_0$ gives rise to a level decomposition of $\frake_{n+1}$ 
for which we denote the subspace
at level $p$ by $\tilde{\scr U_p}$.

The vector spaces $\scr U_p$ and $\tilde{\scr U_p}$ are also modules for $\frake_n$-representations
$R_p$ and $\tilde R_p$, respectively,
given by the adjoint action of the $\frake_n$ subalgebra at level zero.
While $R_1=\tilde R_1$,
the representations $R_2$ and $\tilde R_2$ are different,
contained in 
the symmetric and antisymmetric parts, respectively, of the tensor product $R_1 \times R_1$
(since $\scr {\scr U}_1$ is an odd subspace of the Lie superalgebra $\bor_n$, and $\frake_{n+1}$ is an ordinary Lie algebra).
For $n\leq6$
the Lie algebra $\frake_{n+1}$ is 3-graded with respect to $\frake_n$,
which means that
$\tilde {\scr U}_{p}=0$ for $|p|\geq2$, and thus $\tilde{R}_2$ vanishes.
In the case $n=7$
we instead have a 5-grading,
$\tilde {\scr U}_{p}=0$ for $|p|\geq3$,
where the subspaces $\tilde {\scr U}_{\pm2}$ are one-dimensional,
so that $\tilde{R}_2={\bf 1}$.
For $n=8$, the extended Lie
algebra $\frake_{n+1}=\frake_9$ is the affine extension of $\frake_8$ with infinitely many subspaces $\tilde {\scr U}_p$,
and $\tilde R_p$ is equal to the adjoint representation ${\bf 248}$
of $\frake_8$ for any $p \neq 0$.
(Usually in the definition of $\frake_9$, or any other affine Kac-Moody algebra, a basis element called {\it derivation} is included
at level zero, in addition to those in the definition of $\frake_{n+1}$ above. It can be identified with $h_{-1}$ in the further extension
to $\bor_{n+1}=\bor_9$.)
On the other hand, $\bor_n$ is infinite-dimensional for all $n$, and the dimensions of the infinitely many
(possibly reducible)
representations $R_p$ grow with the level $p\geq1$. 
For any $p$, the representations $R_p$ and $R_{-p}$ (or $\tilde{R}_p$ and $\tilde{R}_{-p}$) are conjugate to each other.

Clearly $R_1$ is an irreducible representation with a lowest weight vector $e_0$ and lowest weight $-\Lambda^1$, where the fundamental weights
$\Lambda^i$
are defined by $(\alpha_i,\Lambda^j)=\de_i{}^j$. 
The dimension of $R_1$ is
$6,10,16,27,56,248$ for $n=3,4,5,6,7,8$, respectively.
The representation
$R_2$ is irreducible for $n\leq7$ with lowest weight $-\Lambda^{n-1}$. For $n=8$, it
decomposes into a direct sum of a representation with lowest weight $-\Lambda^{n-1}$
and an additional singlet.
Tables with the representations $R_p$ for all $n$ and the first few positive levels $p$ 
can be found in for example \cite{Howe:2015hpa},
and an efficient recursive method to compute them for all positive levels was given in
\cite{Cederwall:2015oua}.
In section
\ref{reducibilitet} we will see that 
the sequence of representations $R_p$ for $p\geq1$
is related to the infinite reducibility of the generalized diffeomorphisms in exceptional geometry, as was observed in
\cite{Berman:2012vc}.

Each subspace $\scr U_p$ in the level decomposition of $\scr B_n$ with respect to the gray node
can in turn be further decomposed with respect to any of the other nodes.
Choosing node $n$ (referring to the Dynkin diagram in the beginning of this section),
we then write $\scr U_{\pm1} = \scr U_{\pm1}{}^0 \oplus \scr U_{\pm1}{}^\pm$
where $\scr U_{\pm1}{}^0$ 
is an $n$-dimensional subspace spanned by root vectors for which the roots have zero coefficients corresponding to $\alpha_n$ in the basis of simple roots,
and $\scr U_{\pm1}{}^\pm$ is spanned by root vectors for which
the sign of this coefficient is $\pm1$.
The subalgebra of $\bor_{n}$ generated by $\scr U_{\pm1}{}^0$
is the 3-graded Lie superalgebra
\begin{align} \label{3-grading1}
A(n-1,0) = \sl(n|1)=\scr U_{-1}{}^0 \,\oplus\, \gl(n) \,\oplus\, \scr U_1{}^0,
\end{align}
obtained by removing node $n$ from the Dynkin diagram of $\bor_n$.
In the same way
we write
$\tilde{\scr U}_{\pm1} = \tilde{\scr U}_{\pm1}{}^0 \oplus \tilde{\scr U}_{\pm1}{}^\pm$,
where the subalgebra of $\frake_{n+1}$ generated by $\tilde{\scr U}_{\pm1}{}^0$
is the 3-graded Lie algebra
\begin{align} \label{3-grading0}
\mathfrak{a}_n=\sl(n+1) = \tilde{\scr U}_{-1}{}^0 \,\oplus\, \gl(n) \,\oplus\, \tilde{\scr U}_1{}^0.
\end{align}
The adjoint action of the subalgebra $\gl(n)$ at level zero on $\scr U_1{}^0$ and $\tilde{\scr U}_1{}^0$ is given by the
$n$-dimensional vector (or fundamental) representation.

Thus the restriction of ${\scr U}_{1}$ to ${\scr U}_{1}{}^0$ at level one in $\bor_n$ leads to the restriction of $\frake_n\oplus\mathbb{R}$
to $\mathfrak{gl}(n)$ at level zero, which means reducing exceptional geometry to ordinary geometry (as will be more clear in
the next section).
For this reason, we use indices
\begin{align}
\pzc m,\pzc n,\ldots&=1,2,\ldots,n, & \pzc M,\pzc N,\ldots&=1,2,\ldots,{\rm dim}\, R_1,
\end{align}
and let $E_{\pzc M}$
be a basis of $\scr U_1$
such that
$E_{\pzc m}$
is a basis of $\scr U_1{}^0$.
We then let $F^{\pzc M}$ be a basis of $\scr U_{-1}$ such that
\begin{align}
\la E_{\pzc M}|F^{\pzc N}\ra=
-\la F^{\pzc N}|E_{\pzc M}\ra=\de_{\pzc M}{}^{\pzc N},
\end{align}
which implies that $F^{\pzc m}$
is a basis of $\scr U_{-1}{}^0$.
Considering $\frake_{n+1}$ as a subalgebra of $\bor_{n+1}$ these bases of $\scr U_{\pm1}$ give rise to corresponding bases
$\tilde E_{\pzc M} = [e_{-1},E_{\pzc M}]$ and $\tilde F^{\pzc M} = -[f_{-1},F^{\pzc M}]$ of 
$\tilde{\scr U}_{\pm1}$
such that
\begin{align}
\la \tilde E_{\pzc M}|\tilde F^{\pzc N}\ra=
\la \tilde F^{\pzc N}|\tilde E_{\pzc M}\ra=\de_{\pzc M}{}^{\pzc N}.
\end{align}

\subsection{Commutation relations}

The commutation relations in 
$\bor_n$ and $\frake_{n+1}$
of the elements at level $\pm1$ with each other and with those at level zero
were given in \cite{Palmkvist:2011vz}. We let $t_\alpha$ be a basis of $\frake_n$, raise
the adjoint $\frake_n$ index ${}_\al$ with the inverse of the Killing form, 
(so that
$\la t_\al | t^\be \ra =\de_\al{}^\be$),
and introduce the Cartan elements
\begin{align}
h=(9-n)h_0+(10-n)h_1+\cdots+6h_{n-3}+4h_{n-2}+2h_{n-1}+3h_n
\end{align}
and $\tilde h=h+(9-n)h_{-1}$, which span the orthogonal complements to $\frake_n$ at level zero in 
$\bor_n$ and $\frake_{n+1}$, respectively (so that $[t_\alpha,h]=[t_\alpha,\tilde h]=0$).
Then the relations are
\begin{align}
[E_{\pzc M},F^{\pzc N}]&=(t^\alpha)_{\pzc M}{}^{\pzc N} t_\alpha + \frac1{9-n}\de_{\pzc M}{}^{\pzc N} h,&
[\tilde E_{\pzc M},\tilde F^{\pzc N}]&=(t^\alpha)_{\pzc M}{}^{\pzc N} t_\alpha + \frac1{9-n}\de_{\pzc M}{}^{\pzc N} \tilde h,
\nn\\
[t_\alpha,E_{\pzc M}]&=(t_\alpha)_{\pzc M}{}^{\pzc N} E_{\pzc N}, & 
[t_\alpha,\tilde E_{\pzc M}]&=(t_\alpha)_{\pzc M}{}^{\pzc N}\tilde E_{\pzc N},\nn\\
[t_\alpha,F^{\pzc N}]&=-(t_\alpha)_{\pzc M}{}^{\pzc N} F^{\pzc M}, & 
[t_\alpha,\tilde F^{\pzc N}]&=-(t_\alpha)_{\pzc M}{}^{\pzc N}\tilde F^{\pzc M},\nn\\
[h,E_{\pzc M}]&=-(10-n)E_{\pzc M},  &
[\tilde h,\tilde E_{\pzc M}]&=(8-n)\tilde E_{\pzc M},\nn\\
[h,F^{\pzc N}]&=(10-n)F^{\pzc N},  &
[\tilde h,\tilde F^{\pzc N}]&=-(8-n)\tilde F^{\pzc N}.
\label{borcherds-comm-rel}
\end{align}
Following \cite{Palmkvist:2011vz} (but replacing $g$ there with $\tilde{f}$) we introduce
the $\frake_n$ invariant tensors
\begin{align}
f_{\pzc M}{}^{\pzc N}{}_{\pzc P}{}^{\pzc Q} &= \la [[E_{\pzc M},F^{\pzc N}],E_{\pzc P}]| F^{\pzc Q}\ra, & \tilde f_{\pzc M}{}^{\pzc N}{}_{\pzc P}{}^{\pzc Q} &=
\la [[\tilde E_{\pzc M},\tilde F^{\pzc N}],\tilde E_{\pzc P}]| \tilde F^{\pzc Q}\ra
\end{align}
(which are in fact structure constants of generalized Jordan triple products, as we will see in section \ref{jordan-section})
and their (anti-)symmetrized versions
\begin{align}
f_{\pzc M\,\!\pzc N}{}^{\pzc P\,\!\pzc Q}=\la [E_{\pzc M},E_{\pzc N}] | [F^{\pzc P},F^{\pzc Q}]\ra &=
-2f_{(\pzc N}{}^{\pzc P}{}_{{\pzc M})}{}^{\pzc Q}, \nn\\
\tilde f_{\pzc M\,\!\pzc N}{}^{\pzc P\,\!\pzc Q}=\la [\tilde E_{\pzc M},\tilde E_{\pzc N}] | [\tilde F^{\pzc P},\tilde F^Q]\ra &= 
-2\tilde f_{[\pzc N}{}^{\pzc P}{}_{{\pzc M}]}{}^{\pzc Q}.
\end{align}
We thus have
\begin{align} \label{lilla-f}
[[E_{\pzc M},F^{\pzc N}],E_{\pzc P}]&=f_{\pzc M}{}^{\pzc N}{}_{\pzc P}{}^{\pzc Q} E_{\pzc Q}, &
[[\tilde E_{\pzc M},\tilde F^{\pzc N}],\tilde E_{\pzc P}]&=\tilde f_{\pzc M}{}^{\pzc N}{}_{\pzc P}{}^{\pzc Q} \tilde E_{\pzc Q},\\
\label{lilla-f2}
[[E_{\pzc M},E_{\pzc N}],F^{\pzc P}]&=f_{\pzc M\,\!\pzc N}{}^{\pzc P\,\!\pzc Q} E_{\pzc Q}, &
[[\tilde E_{\pzc M},\tilde E_{\pzc N}],\tilde F^{\pzc P}]&=\tilde f_{\pzc M\,\!\pzc N}{}^{\pzc P\,\!\pzc Q} \tilde E_{\pzc Q},
\end{align}
and from (\ref{borcherds-comm-rel}) we get explicitly
\begin{align}
f_{\pzc M}{}^{\pzc N}{}_{\pzc P}{}^{\pzc Q} &= (t_\alpha)_{\pzc M}{}^{\pzc N} (t^\alpha)_{\pzc P}{}^{\pzc Q}
- \frac{10-n}{9-n}\de_{\pzc M}{}^{\pzc N} \de_{\pzc P}{}^{\pzc Q},\nn\\
\label{originaltripleproduct2}
\tilde{f}_{\pzc M}{}^{\pzc N}{}_{\pzc P}{}^{\pzc Q} &= (t_\alpha)_{\pzc M}{}^{\pzc N} (t^\alpha)_{\pzc P}{}^{\pzc Q} +
\frac{8-n}{9-n}\de_{\pzc M}{}^{\pzc N} \de_{\pzc P}{}^{\pzc Q}.
\end{align}
Restricting the basis elements to $\scr U_{\pm1}{}^0$ and $\tilde{\scr U}_{\pm1}{}^0$ we get
\begin{align}\label{apa}
f_{\pzc m}{}^{\pzc n}{}_{\pzc p}{}^{\pzc q}&=-\de_{\pzc m}{}^{\pzc n}\de_{\pzc p}{}^{\pzc q}+\de_{\pzc p}{}^{\pzc n}\de_{\pzc m}{}^{\pzc q}&
{\tilde f}_{\pzc m}{}^{\pzc n}{}_{\pzc p}{}^{\pzc q}&=\de_{\pzc m}{}^{\pzc n}\de_{\pzc p}{}^{\pzc q}+\de_{\pzc p}{}^{\pzc n}\de_{\pzc m}{}^{\pzc q}.
\end{align}
In accordance with the 3-gradings (\ref{3-grading1}) and (\ref{3-grading0})
we have
\begin{align}
[E_{\pzc m},E_{\pzc n}]=[\tilde{E}_{\pzc m},\tilde{E}_{\pzc n}]=[F^{\pzc m},F^{\pzc n}]=[\tilde{F}^{\pzc m},\tilde{F}^{\pzc n}]=0
\end{align}
and thus
\begin{align} \label{sc-hjalp}
f_{\pzc m\,\!\pzc n}{}^{\pzc M\,\!\pzc N}=\tilde{f}_{\pzc m\,\!\pzc n}{}^{\pzc M\,\!\pzc N}
=f_{\pzc M\,\!\pzc N}{}^{\pzc m\,\!\pzc n}=\tilde{f}_{\pzc M\,\!\pzc N}{}^{\pzc m\,\!\pzc n}=0.
\end{align}
Considering 
$\bor_n$ and $\frake_{n+1}$ as subalgebras of $\bor_{n+1}$ we can generalize (\ref{lilla-f}) to the set of relations
\begin{align}
[[E_{\pzc M},F^{\pzc N}],E_{\pzc P}]&= f_{\pzc M}{}^{\pzc N}{}_{\pzc P}{}^{\pzc Q}E_{\pzc Q},	&
[[\tilde E_{\pzc M},\tilde F^{\pzc N}],\tilde E_{\pzc P}]&=
{\tilde f}_{\pzc M}{}^{\pzc N}{}_{\pzc P}{}^{\pzc Q} \tilde E_{\pzc Q},				
\nn\\
[[E_{\pzc M},F^{\pzc N}],\tilde E_{\pzc P}]&= \de_{\pzc M}{}^{\pzc N}\tilde E_{\pzc P} + f_{\pzc M}{}^{\pzc N}{}_{\pzc P}{}^{\pzc Q}\tilde E_{\pzc Q},
&
[[\tilde E_{\pzc M},\tilde F^{\pzc N}],E_{\pzc P}]&=\de_{\pzc M}{}^{\pzc N}  E_{\pzc P}+ f_{\pzc M}{}^{\pzc N}{}_{\pzc P}{}^{\pzc Q} E_{\pzc Q},			
\nn\\
[[E_{\pzc M},\tilde F^{\pzc N}],E_{\pzc P}]&= 0,&
[[\tilde E_{\pzc M},F^{\pzc N}],\tilde E_{\pzc P}]&= 0,			
\nn\\
[[E_{\pzc M},\tilde F^{\pzc N}],\tilde E_{\pzc P}]&=\de_{\pzc M}{}^{\pzc N} E_{\pzc P},&
[[\tilde E_{\pzc M},F^{\pzc N}],E_{\pzc P}]&=  -\de_{\pzc M}{}^{\pzc N} \tilde E_{\pzc P},
\label{delprodukter}
\end{align}
which 
will be useful in the next section.

\section{The generalized diffeomorphisms}

We will now relate the algebraic concepts introduced in the preceding section to the context of
eleven-dimensional supergravity (or M-theory) compactified to $D=(11-n)$ dimensions.
At each point in the
eleven-dimensional spacetime manifold, the $n$-dimensional 
subspace of the 
tangent space corresponding to the $n$ compactified dimensions
can be considered as a
vector
module of $\gl(n)$, 
which in turn can be extended to a module for the representation $R_1$ of $\frake_n$.
The idea 
is to identify
this $\frake_n$-module with the subspace $\scr U_1{}$,
the $n$-dimensional subspace of $\scr U_1{}$ corresponding to the $n$ compactified dimensions
with $\scr U_1{}^0$, and their coordinate bases with $E_{\pzc M}$ and $E_{\pzc m}$, respectively.
We furthermore ignore the remaining $D$ dimensions, and thus consider
any vector field $V$ as an element in $\scr U_1$, expanded in the coordinate basis as $V=V^{\pzc M} E_{\pzc M}$.

\subsection{The section condition}

We are interested in fields that 
only depend on the $n$ physical coordinates corresponding to the $\scr U_1{}^0$ subspace of
$\scr U_1$. Any such field $A$ thus satisfies
\begin{align} \label{non-kov-sc}
E_{\pzc M} \in \scr U_1{}^+  \qquad&\Rightarrow  \qquad\partial_{\pzc M} A=0.
\end{align}
Because of (\ref{sc-hjalp}) this implies
the {\it section condition}
\begin{align} \label{kov-sc}
f_{\pzc M\,\!\pzc N}{}^{\pzc P\,\!\pzc Q} \,\partial_{\pzc P}\partial_{\pzc Q} A =
f_{\pzc M\,\!\pzc N}{}^{\pzc P\,\!\pzc Q} \,\partial_{\pzc P}A\,\partial_{\pzc Q} B =
\tilde{f}_{\pzc M\,\!\pzc N}{}^{\pzc P\,\!\pzc Q} \,\partial_{\pzc P}A\,\partial_{\pzc Q} B = 0
\end{align}
for any field $A$, or any pair of fields $A$ and $B$.
Following \cite{Berman:2012vc} we write this as
\begin{align} \label{kov-sc2}
f_{\pzc M\,\!\pzc N}{}^{\pzc P\,\!\pzc Q} \,(\partial_{\pzc P}\otimes\partial_{\pzc Q}) =
\tilde f_{\pzc M\,\!\pzc N}{}^{\pzc P\,\!\pzc Q} \,(\partial_{\pzc P}\otimes\partial_{\pzc Q}) = 0,
\end{align}
where $(\partial_{\pzc P}\otimes\partial_{\pzc Q})$ denotes either $\partial_{\pzc P}\partial_{\pzc Q} A$ or
$\partial_{\pzc P}A\,\partial_{\pzc Q} B$.
We can also equivalently replace $f_{\pzc M\,\!\pzc N}{}^{\pzc P\,\!\pzc Q}$ and $\tilde f_{\pzc M\,\!\pzc N}{}^{\pzc P\,\!\pzc Q}$
with the projectors $(\mathbb{P}_2)_{\pzc M\,\!\pzc N}{}^{\pzc P\,\!\pzc Q}$ and $(\tilde{\mathbb{P}}_2)_{\pzc M\,\!\pzc N}{}^{\pzc P\,\!\pzc Q}$
of $R_2$ and $\tilde R_2$,
respectively,
since this only amounts to a rescaling of each irreducible part of these representations.
The section condition then becomes
\begin{align} \label{sc}
(\mathbb{P}_2)_{\pzc M\,\!\pzc N}{}^{\pzc P\,\!\pzc Q} \,(\partial_{\pzc P}\otimes\partial_{\pzc Q}) =
(\tilde{\mathbb{P}}_2)_{\pzc M\,\!\pzc N}{}^{\pzc P\,\!\pzc Q} \,(\partial_{\pzc P}\otimes\partial_{\pzc Q}) = 0.
\end{align}
Unlike the original constraint 
(\ref{non-kov-sc}) the section condition (\ref{kov-sc}) is $\frake_n$-covariant, and
is implied not only by
(\ref{non-kov-sc}) but also 
by solutions
equivalent to (\ref{non-kov-sc}), where $\scr U_1{}^+$ is replaced with different
subspaces of $\scr U_1$
that can be mapped to $\scr U_1{}^+$ by
$\frake_n$-transformations.

In addition there are
solutions where the fields only depend on $(n-1)$ coordinates, corresponding to 
the subspace of $\scr U_1$ 
at level zero in a further decomposition with respect to node $(n-2)$,
spanned by root vectors for which the roots have zero coefficients corresponding to $\alpha_{n-2}$ in the basis of simple roots
(or equivalent solutions obtained by $\frake_n$-transformations). These solutions correspond to
compactification of type IIB supergravity from ten to $(11-n)$ dimensions \cite{Hohm:2013vpa,Hohm:2013uia,Hohm:2014fxa}. 

\subsection{Expressions for the generalized Lie derivative} \label{expressions}

Under a generalized diffeomorphism generated by a vector field $U$, the transformation of another vector field $V$ is given by the 
generalized Lie derivative
\begin{align} \label{genlieder}
\scr L_U V^{\pzc M} &= 
U^{\pzc N}\partial_{\pzc N} V^{\pzc M} - V^{\pzc N}\partial_{\pzc N} U^{\pzc M} +
Y^{\pzc M\,\!\pzc N}{}_{\pzc P\,\!\pzc Q}\partial_{\pzc N} U^{\pzc P} V^{\pzc Q}\nn\\
&= U^{\pzc N}\partial_{\pzc N} V^{\pzc M} + Z^{\pzc M\,\!\pzc N}{}_{\pzc P\,\!\pzc Q}\partial_{\pzc N} U^{\pzc P} V^{\pzc Q},
\end{align}
where
$Y^{\pzc M\,\!\pzc N}{}_{\pzc P\,\!\pzc Q}$ and $Z^{\pzc M\,\!\pzc N}{}_{\pzc P\,\!\pzc Q}=
Y^{\pzc M\,\!\pzc N}{}_{\pzc P\,\!\pzc Q}-\de_{\pzc P}{}^{\pzc M}\de_{\pzc Q}{}^{\pzc N}$ are $\frake_n$-invariant tensors.
These transformations were defined in \cite{Coimbra:2011ky}
and explicitly reconstructed from the $\frake_n$-covariant ansatz above in \cite{Berman:2012vc}.
It was found in \cite{Berman:2012vc} 
that they
close into an algebra
according to
\begin{align}
[\scr L_U, \scr L_V] = \scr L_{\dlb U,V \drb},
\end{align}
where ${\dlb U,V \drb}$ denotes the antisymmetrized generalized Lie derivative,
\begin{align}
\scr L_{\dlb U,V \drb} = \tfrac12(\scr L_U V - \scr L_V U),
\end{align}
if the tensor $Y$
satisfies the identities
\begin{align}
Y^{\pzc M\,\!\pzc N}{}_{\pzc P\,\!\pzc Q}\,\partial_{\pzc M} \otimes \partial_{\pzc N}&=0,\label{y-identitet1}\\
(Y^{\pzc M\,\!\pzc N}{}_{\pzc T\,\!\pzc Q}Y^{\pzc T\,\!\pzc P}{}_{\pzc R\,\!\pzc S}
-Y^{\pzc M\,\!\pzc N}{}_{\pzc R\,\!\pzc S}\de^{\pzc P}{}_{\pzc Q})\partial_{(\pzc N}\otimes\partial_{\pzc P)}&=0,\label{y-identitet2}\\
(Y^{\pzc M\,\!\pzc N}{}_{\pzc T\,\!\pzc Q}Y^{\pzc T\,\!\pzc P}{}_{[\pzc S\,\!\pzc R]}
+2Y^{\pzc M\,\!\pzc N}{}_{[\pzc R|\pzc T|}Y^{\pzc T\,\!\pzc P}{}_{\pzc S]\pzc Q}\qquad\qquad\quad&\nn\\
-Y^{\pzc M\,\!\pzc N}{}_{[\pzc R\,\!\pzc S]}\de^{\pzc P}{}_{\pzc Q}
-2Y^{\pzc M\,\!\pzc N}{}_{[\pzc S|\pzc Q|}\de^{\pzc P}{}_{\pzc R]})\partial_{(\pzc N}\otimes\partial_{\pzc P)}&=0,\label{y-identitet3}\\
(Y^{\pzc M\,\!\pzc N}{}_{\pzc T\,\!\pzc Q}Y^{\pzc T\,\!\pzc P}{}_{(\pzc S\,\!\pzc R)}
+2Y^{\pzc M\,\!\pzc N}{}_{(\pzc R|\pzc T|}Y^{\pzc T\,\!\pzc P}{}_{\pzc S)\pzc Q}\qquad\qquad\quad&\nn\\
-Y^{\pzc M\,\!\pzc N}{}_{(\pzc R\,\!\pzc S)}\de^{\pzc P}{}_{\pzc Q}
-2Y^{\pzc M\,\!\pzc N}{}_{(\pzc S|\pzc Q|}\de^{\pzc P}{}_{\pzc R)})\partial_{[\pzc N}\otimes\partial_{\pzc P]}&=0.\label{y-identitet4}
\end{align}
Up to
the section condition (\ref{sc}), these identities uniquely determine the tensor $Y$, which in \cite{Berman:2012vc} was found to be
\begin{align} \label{y-uttryck}
Y^{\pzc M\,\!\pzc N}{}_{\pzc P\,\!\pzc Q} &= -(t_\al)_{\pzc Q}{}^{\pzc M}(t^\al)_{\pzc P}{}^{\pzc N} 
+ \frac1{9-n}\de_{\pzc Q}{}^{\pzc M}\de_{\pzc P}{}^{\pzc N}+\de_{\pzc P}{}^{\pzc M}\de_{\pzc Q}{}^{\pzc N},
\end{align}
and thus the tensor $Z$ is
\begin{align} \label{z-uttryck}
Z^{\pzc M\,\!\pzc N}{}_{\pzc P\,\!\pzc Q} &= Y^{\pzc M\,\!\pzc N}{}_{\pzc P\,\!\pzc Q}-\de_{\pzc P}{}^{\pzc M}\de_{\pzc Q}{}^{\pzc N}=
-(t_\al)_{\pzc Q}{}^{\pzc M}(t^\al)_{\pzc P}{}^{\pzc N} + \frac1{9-n}\de_{\pzc Q}{}^{\pzc M}\de_{\pzc P}{}^{\pzc N}.
\end{align}
Comparing (\ref{z-uttryck}) with
(\ref{originaltripleproduct2})
we now find that
\begin{align}
f_{\pzc M}{}^{\pzc N}{}_{\pzc P}{}^{\pzc Q} + \tilde{f}_{\pzc M}{}^{\pzc N}{}_{\pzc P}{}^{\pzc Q}
&=-2Z^{\pzc N\,\!\pzc Q}{}_{\pzc P\,\!\pzc M},
\end{align}
which can be inserted in the second term in the second line of (\ref{genlieder}), in order to express the generalized Lie derivative in terms of
$\scr B_n$ and $\frake_{n+1}$.
Although there is no need to simplify the first term
on the right hand side of (\ref{genlieder}), it is interesting to note
that it can be rewritten in a similar way, using
\begin{align}
f_{\pzc M}{}^{\pzc N}{}_{\pzc P}{}^{\pzc Q}- \tilde{f}_{\pzc M}{}^{\pzc N}{}_{\pzc P}{}^{\pzc Q}&=-2\de_{\pzc M}{}^{\pzc N}\de_{\pzc P}{}^{\pzc Q}.
\end{align}
The full expression for the generalized Lie derivative then becomes
\begin{align}
\scr L_U V^{\pzc Q} &= 
-\tfrac12(f_{\pzc M}{}^{\pzc N}{}_{\pzc P}{}^{\pzc Q}
-\tilde{f}_{\pzc M}{}^{\pzc N}{}_{\pzc P}{}^{\pzc Q})U^{\pzc M}\partial_{\pzc N}V^{\pzc P}\nn\\
&\quad\,-\tfrac12(f_{\pzc M}{}^{\pzc N}{}_{\pzc P}{}^{\pzc Q}+\tilde{f}_{\pzc M}{}^{\pzc N}{}_{\pzc P}{}^{\pzc Q})
\partial_{\pzc N}U^{\pzc M}V^{\pzc P}. \label{fg-form}
\end{align}
Note that we get back the ordinary Lie derivative,
\begin{align}
L_U V^{\pzc m}=U^{\pzc n}\partial_{\pzc n}V^{\pzc m}-\partial_{\pzc n}U^{\pzc m}V^{\pzc n},
\end{align}
from (\ref{fg-form}) by restricting $\scr U_1$ and $\tilde{\scr U}_1$ to $\scr U_1{}^+$ and $\tilde{\scr U}_1{}^+$, respectively, and using
(\ref{apa}).

By considering $\scr B_n$ and $\frake_{n+1}$ as subalgebras of $\bor_{n+1}$
it is
possible to obtain an expression
where the components of the vector fields do not appear explicitly.
If we set
$\tilde{V}=[e_{-1},V]=V^{\pzc M} \tilde{E}_{\pzc M}$
for any vector field $V$, then it
follows from
(\ref{delprodukter}) that
\begin{align} \label{bracketform}
\scr L_U V = [[U,\tilde F^{\pzc N}],\partial_{\pzc N}\tilde V]-[[\partial_{\pzc N}\tilde U,\tilde F^{\pzc N}],V].
\end{align}
Applying the adjoint action of $e_{-1}$ to both sides of (\ref{bracketform}), and rewriting the
right hand side using (\ref{delprodukter}), we get the equivalent expression
\begin{align} \label{bracketform2}
\scr L_U \tilde V = -[[\tilde U, F^{\pzc N}],\partial_{\pzc N} V]-[[\partial_{\pzc N} U, F^{\pzc N}],\tilde V],
\end{align}
which turns out to be more useful in analyzing the reducibility of the generalized diffeomorphisms, as we will see in section \ref{reducibilitet}.

The component-free expressions (\ref{bracketform}) and (\ref{bracketform2})
make the reduction of exceptional geometry to ordinary geometry described above more clear. It simply amounts to restricting
the vector fields $U$ and $V$, considered as elements in (the subspace $\scr U_1$ of) the Borcherds superalgebra
$\bor_n$, to the subalgebra $A(n-1,0)=\sl(n|1)$, obtained by removing node $n$ from the Dynkin diagram of $\bor_n$.
Similarly, we obtain the generalized Lie derivative in doubled geometry with T-duality group ${\rm O}(d,d)$,
where $d=n-1,$
from (\ref{bracketform}) by restricting
the vector fields to the subalgebra $D(d,1)=\mathfrak{osp}(2d|2)$
corresponding to removing node $(n-1)$ from the Dynkin diagram of $\bor_n$.

Another advantage of the expressions (\ref{bracketform})--(\ref{bracketform2})
is that the commutator of two generalized Lie derivatives can be computed using the Jacobi identity
in $\scr B_{n+1}$.
In practice it seems however easier to
keep the expression (\ref{genlieder}) and use the identities (\ref{y-identitet1})--(\ref{y-identitet4}),
but as we will see in the next section, these identities can
be derived from the Jacobi identity in $\scr B_{n+1}$, except for half of the identity (\ref{y-identitet2}),
for which some additional information is needed.

\subsection{Closure}
\label{jordan-section}

The closure of the generalized
diffeomorphisms into a Lie algebra
relies on the identities (\ref{y-identitet1})--(\ref{y-identitet4}) for the tensor $Y$,
which we will now derive
by expressing $Y$ in terms of $f$ and $\tilde{f}$. This can be done 
in various ways,
\begin{align}
Y^{\pzc N\,\!\pzc Q}{}_{\pzc P\,\!\pzc M} &= - f_{\pzc M}{}^{\pzc N}{}_{\pzc P}{}^{\pzc Q} + 2\de_{[\pzc P}{}^{\pzc N}\de_{\pzc M]}{}^{\pzc Q} 
= - \tilde{f}_{\pzc M}{}^{\pzc N}{}_{\pzc P}{}^{\pzc Q} + 2\de_{(\pzc P}{}^{\pzc N}\de_{\pzc M)}{}^{\pzc Q}\nn\\
&=-f_{(\pzc M}{}^{\pzc N}{}_{\pzc P)}{}^{\pzc Q}-\tilde{f}_{[\pzc M}{}^{\pzc N}{}_{\pzc P]}{}^{\pzc Q}
=\tfrac12 f_{\pzc P\,\!\pzc M}{}^{\pzc N\,\!\pzc Q}+\tfrac12 \tilde{f}_{\pzc P\,\!\pzc M}{}^{\pzc N\,\!\pzc Q}.
\end{align}
From the last expression it follows that (\ref{y-identitet1}) is equivalent to the section condition (\ref{kov-sc2}).
We will show that this condition, together with the Jacobi identity in $\bor_{n+1}$, implies (\ref{y-identitet3}), (\ref{y-identitet4})
and the part of (\ref{y-identitet2}) symmetric in the indices ${}_{\pzc R}$ and ${}_{\pzc S}$.
To derive the part of (\ref{y-identitet2}) antisymmetric in ${}_{\pzc R}$ and ${}_{\pzc S}$ we also need the fact that
$\tilde{\scr U}_{\pm2}$ is at most one-dimensional for $n\leq7$.

Let $\hat{\scr U}_1={\scr U}_1\oplus\tilde{\scr U}_1$ be
the level-one subspace of $\bor_{n+1}$
in the level decomposition with respect to node $0$ (the innermost of the two gray nodes) in the Dynkin diagram.
Let $\tau$ be a vector space automorphism of $\scr B_{n+1}$ such that
$\tau([x,y])=[\tau(y),\tau(x)]$,
preserving the $\mathbb{Z}_2$-degree but reversing the $\mathbb{Z}$-degree, so that
$\tau(\scr U_{\pm 1})=\scr U_{\mp1}$ and $\tau(\tilde{\scr U}_{\pm 1})=\tilde{\scr U}_{\mp1}$.
As a consequence of the Jacobi identity in $\bor_{n+1}$, the triple product
\begin{align} \label{trippelprodukt}
\hat{\scr U}_1 \times \hat{\scr U}_1 \times \hat{\scr U}_1 &\to \hat{\scr U}_1,& (x,y,z) &\mapsto (xyz)\equiv[[x,\tau(y)],z],
\end{align} 
then satisfies the identity
\begin{align} \label{super-fund-id}
(uv(xyz))-(-1)^\sigma(xy(uvz))=((uvx)yz)-(-1)^\sigma(x(vuy)z),
\end{align}
where 
$\sigma = {(|u|+|v|)(|x|+|y|)}$, denoting the $\mathbb{Z}_2$-degree of any element $z$ by $|z|$. Indeed, the Jacobi identity turns the left hand side of
(\ref{super-fund-id}) into
\begin{align}
\Big[\big[[u,\tau(v)],[x,\tau(y)]\big],z\Big],
\end{align}
which after using the Jacobi identity once again becomes
\begin{align}
\bigg[\Big[\big[[u,\tau(v)],x\big],\tau(y)\Big],z\bigg]-(-1)^{|x||y|}\bigg[\Big[\big[[u,\tau(v)],\tau(y)\big],x\Big],z\bigg],
\end{align}
where the last term is equal to
\begin{align}
-(-1)^\sigma \bigg[\Big[x,\tau\Big(\big[[v,\tau(u)],y\big]\Big)\Big],z\bigg].
\end{align}

We choose the linear map $\tau$ to be given by $\tau(E_{\pzc M})=F^{\pzc M}$ and
$\tau(\tilde{E}_{\pzc M})=\tilde{F}^{\pzc M}$.
According to (\ref{lilla-f}) we then have
\begin{align} \label{hatt-trippel0}
(E_{\pzc M}E_{\pzc N}E_{\pzc P})&=f_{\pzc M}{}^{\pzc N}{}_{\pzc P}{}^{\pzc Q} E_{\pzc Q}, &
(\tilde E_{\pzc M}\tilde E_{\pzc N}\tilde E_{\pzc P})&=\tilde f_{\pzc M}{}^{\pzc N}{}_{\pzc P}{}^{\pzc Q} \tilde E_{\pzc Q},
\end{align}
for the basis elements ${E}_{\pzc M}$ and $\tilde{E}_{\pzc M}$ of ${\scr U}_1$ and $\tilde{\scr U}_1$,
respectively, and thus these subspaces of $\hat{\scr U}_1$
close under the triple product (\ref{trippelprodukt}).
Since they are homogeneous with respect to the $\mathbb{Z}_2$-grading, the identity (\ref{super-fund-id})
on these subspaces becomes 
\begin{align} \label{fund-id}
(xy(uvz))-(uv(xyz))=((xyu)vz)-(x(vuy)z),
\end{align}
which means that they satisfy the definition of a {\it generalized Jordan triple system} \cite{Kantor3.5}.
For $E_{\pzc M}$ the identity (\ref{fund-id})
can be written in component form as
\begin{align} \label{fund-id-compform}
{f}_{\pzc M}{}^{\pzc N}{}_{\pzc S}{}^{\pzc T}{f}_{\pzc P}{}^{\pzc Q}{}_{\pzc R}{}^{\pzc S} -
{f}_{\pzc P}{}^{\pzc Q}{}_{\pzc S}{}^{\pzc T}{f}_{\pzc M}{}^{\pzc N}{}_{\pzc R}{}^{\pzc S} =
{f}_{\pzc M}{}^{\pzc N}{}_{\pzc P}{}^{\pzc S}{f}_{\pzc S}{}^{\pzc Q}{}_{\pzc R}{}^{\pzc T} -
{f}_{\pzc P}{}^{\pzc S}{}_{\pzc R}{}^{\pzc T}{f}_{\pzc M}{}^{\pzc N}{}_{\pzc S}{}^{\pzc Q},
\end{align}
and for
$\tilde{E}_{\pzc M}$ the same identity holds with $f$ replaced with $\tilde{f}$. 

It follows from (\ref{delprodukter}) that also the subspace of $\hat{\scr U}_1$ spanned by all
linear combinations $\hat{E}_{\pzc M}={E}_{\pzc M}+\tilde{E}_{\pzc M}$ closes under the triple product 
(\ref{trippelprodukt}), which for these basis elements is
\begin{align} \label{hatt-trippel}
({\hat E}_{\pzc M}{\hat E}_{\pzc N}{\hat E}_{\pzc P})=({f}_{\pzc M}{}^{\pzc N}{}_{\pzc P}{}^{\pzc Q}
+ \tilde{f}_{\pzc M}{}^{\pzc N}{}_{\pzc P}{}^{\pzc Q}){\hat E}_{\pzc Q}=-2 Z^{\pzc N\,\!\pzc Q}{}_{\pzc P\,\!\pzc M} {\hat E}_{\pzc Q}
\end{align}
and the component form of the identity (\ref{super-fund-id}) for this subspace 
is again given by (\ref{fund-id-compform}), but now with 
$f_{\pzc M}{}^{\pzc N}{}_{\pzc P}{}^{\pzc Q}$
replaced with $Z^{\pzc N\,\!\pzc Q}{}_{\pzc P\,\!\pzc M}$, that is
\begin{align} \label{z-fundid}
Z^{\pzc N\,\!\pzc T}{}_{\pzc S\,\!\pzc M}Z^{\pzc Q\,\!\pzc S}{}_{\pzc R\,\!\pzc P}-Z^{\pzc Q\,\!\pzc T}{}_{\pzc S\,\!\pzc P}Z^{\pzc N\,\!\pzc S}{}_{\pzc R\,\!\pzc M}=
Z^{\pzc N\,\!\pzc S}{}_{\pzc P\,\!\pzc M}Z^{\pzc Q\,\!\pzc T}{}_{\pzc R\,\!\pzc S}-Z^{\pzc S\,\!\pzc T}{}_{\pzc R\,\!\pzc P}Z^{\pzc N\,\!\pzc Q}{}_{\pzc S\,\!\pzc M}.
\end{align}
After symmetrizing
(\ref{z-fundid}) in the indices ${}^{\pzc N}$ and ${}^{\pzc Q}$
the left hand side is antisymmetric in the indices 
${}_{\pzc M}$ and ${}_{\pzc P}$, so the right hand side must be
antisymmetric in ${}_{\pzc M}$ and ${}_{\pzc P}$ as well. Likewise,
after antisymmetrizing (\ref{z-fundid}) in ${}^{\pzc N}$ and ${}^{\pzc Q}$
the
right hand side must be
symmetric in ${}_{\pzc M}$ and ${}_{\pzc P}$.
Thus (\ref{z-fundid}) is equivalent to the set of identities
\begin{align} \label{z-fundid2}
2Z^{(\pzc N|\pzc T}{}_{\pzc S[\pzc M|}Z^{|\pzc Q)\pzc S}{}_{\pzc R|\pzc P]}
-Z^{(\pzc N|\pzc S}{}_{[\pzc P\,\!\pzc M]}Z^{|\pzc Q)\pzc T}{}_{\pzc R\,\!\pzc S}+Z^{\pzc S\,\!\pzc T}{}_{\pzc R[\pzc P|}Z^{(\pzc N\,\!\pzc Q)}{}_{\pzc S|\pzc M]}&=0,\nn\\
2Z^{[\pzc N|\pzc T}{}_{\pzc S(\pzc M|}Z^{|\pzc Q]\pzc S}{}_{\pzc R|\pzc P)}
-Z^{[\pzc N|\pzc S}{}_{(\pzc P\,\!\pzc M)}Z^{|\pzc Q]\pzc T}{}_{\pzc R\,\!\pzc S}+Z^{\pzc S\,\!\pzc T}{}_{\pzc R(\pzc P|}Z^{[\pzc N\,\!\pzc Q]}{}_{\pzc S|\pzc M)}&=0,\nn\\
Z^{(\pzc N|\pzc S}{}_{(\pzc P\,\!\pzc M)}Z^{|\pzc Q)\pzc T}{}_{\pzc R\,\!\pzc S}-Z^{\pzc S\,\!\pzc T}{}_{\pzc R(\pzc P|}Z^{(\pzc N\,\!\pzc Q)}{}_{\pzc S|\pzc M)}&=0,\nn\\
Z^{[\pzc N|\pzc S}{}_{[\pzc P\,\!\pzc M]}Z^{|\pzc Q]\pzc T}{}_{\pzc R\,\!\pzc S}-Z^{\pzc S\,\!\pzc T}{}_{\pzc R[\pzc P|}Z^{[\pzc N\,\!\pzc Q]}{}_{\pzc S|\pzc M]}&=0,
\end{align}
which, using $Z^{\pzc M\,\!\pzc N}{}_{\pzc P\,\!\pzc Q} = Y^{\pzc M\,\!\pzc N}{}_{\pzc P\,\!\pzc Q}-\de_{\pzc P}{}^{\pzc M}\de_{\pzc Q}{}^{\pzc N}$,
can be written
\begin{align}
Y^{\pzc S\,\!\pzc T}{}_{\pzc R[\pzc M}Y^{(\pzc N\,\!\pzc Q)}{}_{\pzc P]\pzc S}+\de^{\pzc T}{}_{[\pzc M}Y^{(\pzc N\,\!\pzc Q)}{}_{\pzc P]\pzc R}&=
Y^{\pzc T(\pzc N}{}_{\pzc S\,\!\pzc R}Y^{|\pzc S|\pzc Q)}{}_{[\pzc P\,\!\pzc M]}+2Y^{\pzc T(\pzc N}{}_{[\pzc M|\pzc S|}Y^{|\pzc S|\pzc Q)}{}_{\pzc P]\pzc R}\nn\\
&\quad\,-Y^{\pzc T(\pzc N}{}_{[\pzc M \,\!\pzc P]}\de^{\pzc Q)}{}_{\pzc R}-2Y^{\pzc T(\pzc N}{}_{[\pzc P|\pzc R|}\de^{\pzc Q)}{}_{\pzc M]},
\label{y-fundid1}\\
Y^{\pzc S\,\!\pzc T}{}_{\pzc R(\pzc M}Y^{[\pzc N\,\!\pzc Q]}{}_{\pzc P)\pzc S}+\de^{\pzc T}{}_{(\pzc M}Y^{[\pzc N\,\!\pzc Q]}{}_{\pzc P)\pzc R}&=
Y^{\pzc T[\pzc N}{}_{\pzc S\,\!\pzc R}Y^{|\pzc S|\pzc Q]}{}_{(\pzc P\,\!\pzc M)}+2Y^{\pzc T[\pzc N}{}_{(\pzc M|\pzc S|}Y^{|\pzc S|\pzc Q]}{}_{\pzc P)\pzc R}\nn\\
&\quad\,-Y^{\pzc T[\pzc N}{}_{(\pzc M\,\!\pzc P)}\de^{\pzc Q]}{}_{\pzc R}-2Y^{\pzc T[\pzc N}{}_{(\pzc P|\pzc R|}\de^{\pzc Q]}{}_{\pzc M)},
\label{y-fundid2}\\
Y^{\pzc S\,\!\pzc T}{}_{\pzc R(\pzc M}Y^{(\pzc N\,\!\pzc Q)}{}_{\pzc P)\pzc S}-\de^{\pzc T}{}_{(\pzc M}Y^{(\pzc N\,\!\pzc Q)}{}_{\pzc P)\pzc R}&=
Y^{\pzc T(\pzc N}{}_{\pzc S\,\!\pzc R}Y^{|\pzc S|\pzc Q)}{}_{(\pzc P\,\!\pzc M)}-Y^{\pzc T(\pzc N}{}_{(\pzc M\,\!\pzc P)}\de^{\pzc Q)}{}_{\pzc R},
\label{y-fundid3}\\
Y^{\pzc S\,\!\pzc T}{}_{\pzc R[\pzc M}Y^{[\pzc N\,\!\pzc Q]}{}_{\pzc P]\pzc S}-\de^{\pzc T}{}_{[\pzc M}Y^{[\pzc N\,\!\pzc Q]}{}_{\pzc P]\pzc R}&=
Y^{\pzc T[\pzc N}{}_{\pzc S\,\!\pzc R}Y^{|\pzc S|\pzc Q]}{}_{[\pzc P\,\!\pzc M]}-Y^{\pzc T[\pzc N}{}_{[\pzc M\,\!\pzc P]}\de^{\pzc Q]}{}_{\pzc R}.
\label{y-fundid4}
\end{align}
Contracting
(\ref{y-fundid1})--(\ref{y-fundid3})
with $\partial_{\pzc N} \otimes \partial_{\pzc Q}$ and using (\ref{y-identitet1}) the left hand sides vanishes and we get,
respectively, 
(\ref{y-identitet3}), (\ref{y-identitet4}) and 
half of (\ref{y-identitet2}),
namely the part
symmetric in the indices ${}_{\pzc R}$ and ${}_{\pzc S}$ (here ${}_{\pzc P}$ and ${}_{\pzc M}$). Doing the same 
with (\ref{y-fundid4})
gives
the additional identity
\begin{align}
(Y^{\pzc M\,\!\pzc N}{}_{\pzc T\,\!\pzc Q}Y^{\pzc T\,\!\pzc P}{}_{[\pzc R\,\!\pzc S]}-
Y^{\pzc M\,\!\pzc N}{}_{[\pzc S\,\!\pzc R]}\de^{\pzc P}{}_{\pzc Q})\partial_{[\pzc N} \otimes \partial_{\pzc P]}&=0,
\end{align}
which is not needed for the closure of the generalized diffeomorphisms. (However, it is in fact needed for their covariance,
${\scr L}_U {\scr L}_V = {\scr L}_{{\scr L}_U V} + {\scr L}_V {\scr L}_U$.)
The remaining antisymmetric part of (\ref{y-identitet2}) reads
\begin{align} \label{remaining}
(Y^{\pzc M\,\!\pzc N}{}_{\pzc T\,\!\pzc Q}Y^{\pzc T\,\!\pzc P}{}_{[\pzc R\,\!\pzc S]}-
Y^{\pzc M\,\!\pzc N}{}_{[\pzc R\,\!\pzc S]}\de^{\pzc P}{}_{\pzc Q})\partial_{(\pzc N} \otimes \partial_{\pzc P)}&=0
\end{align}
and is trivially satisfied for $n\leq6$ since $Y$ then is symmetric in the lower (and upper) indices. It still holds for
$n=7$ but fails for $n=8$. This can be understood from the $\mathbb{Z}$-grading
of $\frake_{n+1}$ with respect to $\frake_{n}$.
For $n\leq6$ this is a 3-grading,
\begin{align}
[\tilde E_{\pzc M},\tilde E_{\pzc N}]=[\tilde F^{\pzc M},\tilde F^{\pzc N}]=0,
\end{align}
which is equivalent to the symmetry of $Y$ in its lower (and upper) indices, since
\begin{align}
Y^{\pzc M\,\!\pzc N}{}_{[\pzc P\,\!\pzc Q]}&=
\tfrac12 \tilde f_{\pzc P\pzc Q}{}^{\pzc M\pzc N}=
\tfrac12 \la [\tilde E_{\pzc P},\tilde E_{\pzc Q}] | [\tilde F^{\pzc M},\tilde F^{\pzc N}] \ra,
\end{align}
while for $n=7$ 
we have
a 5-grading of $\frake_{8}$ with respect to $\frake_{7}$, where the subspaces $\tilde{\scr U}_{\pm2}$ are 
one-dimensional.
This means that
$[\tilde E_{\pzc M},[\tilde F^{\pzc N},[\tilde E_{\pzc P},\tilde E_{\pzc Q}]]]$
must be proportional to $\de_{\pzc M}{}^{\pzc N} [\tilde E_{\pzc P},\tilde E_{\pzc Q}]$, and it is easy to check that this proportionality in fact is an equality. From this equality,
and the 5-grading,
\begin{align}
[\tilde E_{\pzc M},[\tilde E_{\pzc N},\tilde E_{\pzc P}]]=[\tilde F^{\pzc M},[\tilde F^{\pzc N},\tilde F^{\pzc P}]]=0,
\end{align}
we get the identities
\begin{align}
2 \tilde{f}_{\pzc N}{}^{\pzc M}{}_{[\pzc S}{}^{\pzc R}\tilde{f}_{\pzc T]\pzc R}{}^{\pzc P\,\!\pzc Q}=
2 \tilde{f}_{\pzc N}{}^{\pzc M}{}_{\pzc R}{}^{[\pzc P}\tilde{f}_{\pzc S\,\!\pzc T}{}^{\pzc Q]\pzc R}&=
\tilde{f}_{\pzc N\,\!\pzc R}{}^{\pzc P\,\!\pzc Q}\tilde{f}_{\pzc S\,\!\pzc T}{}^{\pzc M\,\!\pzc R}=
-\de_{\pzc N}{}^{\pzc M}\tilde{f}_{\pzc S\,\!\pzc T}{}^{\pzc P\,\!\pzc Q}.
\end{align}
Expressed in $Y$ the last two equations become
\begin{align}
2\de^{[\pzc P}{}_{\pzc N} Y^{\pzc Q]\pzc M}{}_{[\pzc T\,\!\pzc S]}
-2Y^{\pzc M[\pzc P}{}_{\pzc R\,\!\pzc N}Y^{\pzc Q]\pzc R}{}_{[\pzc T\,\!\pzc S]}
=2Y^{[\pzc P\,\!\pzc Q]}{}_{\pzc R\,\!\pzc N}Y^{\pzc M\,\!\pzc R}{}_{[\pzc T\,\!\pzc S]}=
\de^{\pzc M}{}_{\pzc N}Y^{\pzc P\,\!\pzc Q}{}_{[\pzc T\,\!\pzc S]}.
\end{align}
For the antisymmetric part of (\ref{y-identitet2}) we now get
\begin{align}
Y^{\pzc M\,\!\pzc N}{}_{\pzc T\,\!\pzc Q}Y^{\pzc T\,\!\pzc P}{}_{[\pzc R\,\!\pzc S]}
-Y^{\pzc M\,\!\pzc N}{}_{[\pzc R\,\!\pzc S]}\de^{\pzc P}{}_{\pzc Q}&=
-\de^{\pzc P}{}_{\pzc Q}Y^{\pzc M\,\!\pzc N}{}_{[\pzc R\,\!\pzc S]}-Y^{\pzc N\,\!\pzc M}{}_{\pzc T\,\!\pzc Q}Y^{\pzc P\,\!\pzc T}{}_{[\pzc R\,\!\pzc S]}\nn\\
&\quad\,-2Y^{[\pzc M\,\!\pzc N]}{}_{\pzc T\,\!\pzc Q}Y^{\pzc P\,\!\pzc T}{}_{[\pzc R\,\!\pzc S]}\nn\\
&=2(\de^{[\pzc M}{}_{\pzc Q}Y^{\pzc P]\pzc N}{}_{[\pzc R\,\!\pzc S]}-Y^{\pzc N[\pzc M}{}_{\pzc T\,\!\pzc Q}Y^{\pzc P]\pzc T}{}_{[\pzc R\,\!\pzc S]})\nn\\
&\quad\,-\de^{\pzc M}{}_{\pzc Q}Y^{\pzc P\,\!\pzc N}{}_{[\pzc R\,\!\pzc S]}
-Y^{\pzc N\,\!\pzc P}{}_{\pzc T\,\!\pzc Q}Y^{\pzc M\,\!\pzc T}{}_{[\pzc R\,\!\pzc S]}\nn\\
&\quad\,-\de^{\pzc P}{}_{\pzc Q} Y^{\pzc M\,\!\pzc N}{}_{[\pzc R\,\!\pzc S]}\nn\\
&=-\de^{\pzc M}{}_{\pzc Q}Y^{\pzc P\,\!\pzc N}{}_{[\pzc R\,\!\pzc S]}-Y^{\pzc N\,\!\pzc P}{}_{\pzc T\,\!\pzc Q}Y^{\pzc M\,\!\pzc T}{}_{[\pzc R\,\!\pzc S]}\nn\\
&\quad\,-\de^{\pzc P}{}_{\pzc Q} Y^{\pzc M\,\!\pzc N}{}_{[\pzc R\,\!\pzc S]}+\de^{\pzc N}{}_{\pzc Q} Y^{\pzc M\,\!\pzc P}{}_{[\pzc R\,\!\pzc S]},
\end{align}
which after contraction with 
$\partial_{(\pzc N} \otimes \partial_{\pzc P)}$ and using (\ref{y-identitet1}) gives (\ref{remaining}).

\subsection{Reducibility} \label{reducibilitet}

We have seen that
the possible transformations of a vector field $V$ under generalized diffeomorphisms are parametrized by vector fields $U$, which, as well as
$V$,
can be considered as elements in the subspace
$\scr U_1$ of $\scr B_n$.
However, the correspondence between all possible transformations of $V$ and all elements $U$ in $\scr U_1$ is not one-to-one.
If $U$
is given by
$U=\partial_{\pzc M} [U',F^{\pzc M}]$
for some $U'$ in $\scr U_2$,
then the Jacobi identity gives
\begin{align}
[\partial_{\pzc M}[U,F^{\pzc M}],\tilde V] = 
[\partial_{\pzc M}\partial_{\pzc N} [[U',F^{\pzc N}],F^{\pzc M}],\tilde V]=
\tfrac12 [\partial_{\pzc M}\partial_{\pzc N} [U',[F^{\pzc N},F^{\pzc M}]],\tilde V]
\end{align}
for the second term in (\ref{bracketform2}), or (\ref{bracketform2igen}) below.
This vanishes by the section condition since $[F^{\pzc N},F^{\pzc M}]$ belongs to $\scr U_{-2}$,
and thus project $\partial_{\pzc M}\partial_{\pzc N}$ on $R_2$.
Similarly we get
\begin{align}
[[\tilde U,F^{\pzc M}],\partial_{\pzc M} V] = [\partial_{\pzc N} [[\tilde U',F^{\pzc N}],F^{\pzc M}],\partial_{\pzc M}V]
=\tfrac12 [\partial_{\pzc N} [\tilde U',[F^{\pzc N},F^{\pzc M}]],\partial_{\pzc M}V]
\end{align}
for the first term (the transport term),
which also vanishes. Thus any element $U$ in $\scr U_1$ 
given by
$U=\partial_{\pzc M} [U',F^{\pzc M}]$
for some $U'$ in $\scr U_2$ generates a zero transformation.
However, the correspondence between elements in $\scr U_1$ that generate zero transformations 
and general elements in $\scr U_2$ is not one-to-one either,
since $\partial_{\pzc M} [U',F^{\pzc M}]=0$ if $U'$ is given by $U'=\partial_{\pzc M} [U'',F^{\pzc M}]$
for some $U''$ in $\scr U_3$.
Continuing in this way the naive counting for the effective number of parameters is given by the 
alternating sum
\begin{align} \label{altsum}
{\rm dim }\,\scr U_1 - {\rm dim}\,\scr U_2 + {\rm dim }\,\scr U_3 -{\rm dim }\,\scr U_4+\cdots
\end{align}
which is highly divergent since the dimensions of 
$\scr U_p$ increase with the level $p$.
In \cite{Berman:2012vc} it was shown, using a corresponding partition function,
that the alternating sum (\ref{altsum}) can be regularized to give the correct number
coming 
from the decomposition of generalized diffeomorphisms into ordinary diffeomorphisms, 2- and 5-form gauge transformations, and
dual diffeomorphisms. The corresponding 
partition function is related to the denominator formula for $\bor_n$, as shown in \cite{Cederwall:2015oua}.

Introducing an operator $\partial = ({\rm ad}\,F^{\pzc M})\partial_{\pzc M}$ 
the alternating sum (\ref{altsum}) can be viewed as expressing
the homology of the chain complex
\begin{align} \label{altsum2}
\scr U_1 \stackrel{\partial}{\longleftarrow} \scr U_2 \stackrel{\partial}{\longleftarrow} \scr U_3 \stackrel{\partial}{\longleftarrow}
\scr U_4 \stackrel{\partial}{\longleftarrow} \cdots
\end{align}
where the nilpotency $\partial^2=0$ follows from the section condition, again by the Jacobi identity.
In \cite{Cederwall:2013naa} it was shown that the derivative in the operator $\partial$ from $\scr U_{p+1}$
to $\scr U_{p}$ is covariant only for $1\leq p\leq 7-n$.

The infinite reducibility of the generalized diffeomorphisms in exceptional geometry is a qualitative difference compared to
ordinary and doubled geometry. As explained in section \ref{expressions}, these cases 
can be obtained from the exceptional one by restricting $\bor_{n}$ to 
the subalgebras $A(n-1,0)=\sl(n|1)$ and $D(n-1,1)=\mathfrak{osp}(2n-2|2)$, obtained by removing node
$n$ and node $(n-1)$ from the Dynkin diagram of $\bor_{n}$, respectively.
With respect to the gray node, $A(n-1,0)$ is 3-graded and $D(n-1,1)$ is 5-graded, which implies no reducibility in
ordinary geometry and a first-order finite reducibility in double geometry.

\section{Discussion} \label{conclusion}

In this paper we have
studied generalized diffeomorphisms in exceptional geometry with U-duality group $E_{n(n)}$ for $n\leq7$. 
By considering any vector field $V$ as an element in the level-one subspace $\scr U_1$ of the Borcherds superalgebra $\bor_n$,
which in turn is considered as a subalgebra of $\bor_{n+1}$, we have found that the 
generalized Lie derivative of $V$ parametrized by another vector field $U$
can be written
\begin{align} \label{bracketformigen}
\scr L_U V &= [[U,\tilde F^{\pzc N}],\partial_{\pzc N}\tilde V]-[[\partial_{\pzc N}\tilde U,\tilde F^{\pzc N}],V],
\end{align}
or equivalently
\begin{align} 
\scr L_U \tilde V &= -[[\tilde U, F^{\pzc N}],\partial_{\pzc N} V]-[[\partial_{\pzc N} U, F^{\pzc N}],\tilde V],\label{bracketform2igen}
\end{align}
where the tilde on $U$ and $V$ denotes the adjoint action of
$e_{-1}$, and $\tilde{F}^{\pzc M}=-[f_{-1},{F}^{\pzc M}]$.
It would of course be interesting to 
include also other aspects of exceptional geometry in this framework,
for example
concepts of connection,
torsion of curvature
which could be needed for an extension to the case $n=8$ (and beyond).
The transformations (\ref{bracketformigen}) and (\ref{bracketform2igen}) can be defined in precisely the same way for $n=8$ as for $n\leq7$, but 
then they fail to close, as was noted in \cite{Berman:2012vc}. We have shown in this paper that the failure
can be understood from the $\mathbb{Z}$-grading of $\frake_9$ which respect to $\frake_8$.
With the notation in \cite{Cederwall:2015ica} we only get $\ring{\scr L_U} V$ from the right hand side of
(\ref{bracketformigen}) and we have to supplement this expression with an additional term, involving an additional `section-projected' parameter field,
in order to obtain closed and covariant transformations \cite{Hohm:2014fxa,Rosabal:2014rga,Cederwall:2015ica}.

An advantage of the
expressions (\ref{bracketformigen}) and
(\ref{bracketform2igen}) is their universality. They are at the same time valid not only for any $D \geq 4$, but also for ordinary geometry and doubled
geometry as well as for exceptional geometry. The cases of ordinary and doubled geometry
can be obtained from the exceptional case by simply restricting $\bor_{n+1}$ to subalgebras, corresponding
to removing nodes from the Dynkin diagram,
without changing the expression for the generalized Lie derivative.
However, it seems unsatisfactory from the universality point of view
that different algebras $\bor_{n+1}$ are needed for different $n$. This problem can probably be solved by
taking into account also the $D$ external dimensions and adding $D-1$ white nodes to the 
$\bor_{n+1}$ diagram
so that the additional white nodes form a Dynkin diagram of $\mathfrak{a}_{D-1}=\sl(D)$, connected with a line 
from one of its end nodes to the outermost gray node in the
$\bor_{n+1}$ diagram. The different Dynkin 
diagrams
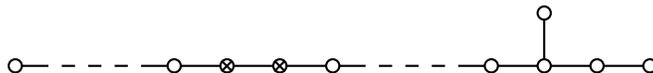
\begin{center}
\begin{picture}(245,25)(-62.5,-2.5)
\thicklines
\put(2.5,0){\line(1,0){15}}
\put(-60,0){\circle{5}}
\put(0,0){\circle{5}}
\put(-57.5,0){\line(1,0){10}}
\put(-42.5,0){\line(1,0){5}}
\put(-32.5,0){\line(1,0){5}}
\put(-22.5,0){\line(1,0){5}}
\put(-12.5,0){\line(1,0){10}}
\put(18,-2){\line(1,1){4}}
\put(22,-2){\line(-1,1){4}}
\put(38,-2){\line(1,1){4}}
\put(42,-2){\line(-1,1){4}}
\put(20,0){\circle{5}}
\put(40,0){\circle{5}}
\put(60,0){\circle{5}}
\put(120,0){\circle{5}}
\put(140,0){\circle{5}}
\put(160,0){\circle{5}}
\put(180,0){\circle{5}}
\put(140,20){\circle{5}}
\put(22.5,0){\line(1,0){15}}
\put(42.5,0){\line(1,0){15}}
\put(62.5,0){\line(1,0){10}}
\put(77.5,0){\line(1,0){5}}
\put(87.5,0){\line(1,0){5}}
\put(97.5,0){\line(1,0){5}}
\put(107.5,0){\line(1,0){10}}
\put(122.5,0){\line(1,0){15}}
\put(142.5,0){\line(1,0){15}}
\put(162.5,0){\line(1,0){15}}
\put(140,2.5){\line(0,1){15}}
\end{picture}
\end{center}
obtained in this way for different $D$ are in fact equivalent in the sense that they describe the same 
Borcherds superalgebra $\scr B_{11}$.
By a series of odd reflections a `distinguished' Dynkin diagram can be reached, with only one gray node connected to the 
Dynkin diagram of
$\frake_{11}$.
If our results can be extended accordingly they could be related to the
$\frake_{11}$ approach in \cite{Berman:2011jh,West:2011mm,West:2012qz,West:2014eza}.

The partial derivatives in (\ref{bracketform2igen}) appear contracted with the basis elements $F^{\pzc M}$ of $\scr U_{-1}$, and as a consequence,
the section condition can be relaxed if the general basis elements $F^{\pzc M}$ 
are replaced by
elements $\Phi^{\pzc M}$ that are not linearly independent, but span a subspace of $\scr U_{-1}$
(equivalent to $\scr U_{-1}{}^0$ for the solutions corresponding to
eleven-dimensional supergravity)
which is `isotropic' in the sense that
$[\Phi^{\pzc M},\Phi^{\pzc N}]=0$ for any pair of elements $\Phi^{\pzc M}$ and $\Phi^{\pzc N}$.
This alternative version of the section condition
might be easier to handle than the original one, and
might be especially useful in further work connecting exceptional geometry
to gauged supergravity (possibly related to the previous work \cite{Aldazabal:2013mya,Aldazabal:2013via,Hohm:2014qga}),
since the truncation of $\bor_{n+1}$ given by the 
projection of $F^{\pzc M}$ on the elements $\Phi^{\pzc M}$ satisfying $[\Phi^{\pzc M},\Phi^{\pzc N}]=0$
is similar to the truncation of the tensor hierarchy algebra \cite{Palmkvist:2013vya},
where the subspace at level $-1$ is spanned by the embedding tensor 
$\Theta$, satisfying the quadratic constraint
$[\Theta,\Theta]=0$.\\

\noindent\underbar{Acknowledgements:}
I would like to thank Martin Cederwall, Sebastian Guttenberg, Axel Kleinschmidt and Ergin Sezgin for valuable discussions, and the
referee for making me aware of a few typos.
The work is supported in part by NSF grants PHY-1214344 and PHY-1521099, and by the
George P.\ and Cynthia Woods Mitchell Institute for Fundamental Physics and Astronomy.

\bibliographystyle{utphysmod2}

\begin{thebibliography}{10}

\bibitem{Hull:2007zu}
C.~Hull,  {\em {Generalised geometry for M-Theory}}, JHEP {\bf 0707}, 079
  (2007)
[\href{http://www.arXiv.org/abs/hep-th/0701203}{{\tt hep-th/0701203}}].

\bibitem{Pacheco:2008ps}
P.~P. Pacheco and D.~Waldram,  {\em {M-theory, exceptional generalised geometry
  and superpotentials}}, JHEP {\bf 0809}, 123 (2008)
[\href{http://www.arXiv.org/abs/0804.1362}{{\tt 0804.1362}}].

\bibitem{Hillmann:2009ci}
C.~Hillmann,  {\em {Generalized $E_{7(7)}$ coset dynamics and $D=11$ supergravity}},
  JHEP {\bf 0903}, 135 (2009)
[\href{http://www.arXiv.org/abs/0901.1581}{{\tt 0901.1581}}].

\bibitem{Berman:2010is}
D.~S. Berman and M.~J. Perry,  {\em {Generalized geometry and M-theory}}, JHEP
  {\bf 1106}, 074 (2011)
[\href{http://www.arXiv.org/abs/1008.1763}{{\tt 1008.1763}}].

\bibitem{Berman:2011pe}
D.~S. Berman, H.~Godazgar and M.~J. Perry,  {\em {${\rm SO}(5,5)$ duality in M-theory
  and generalized geometry}}, Phys.\ Lett.\ {\bf B700}, 65--67 (2011)
[\href{http://www.arXiv.org/abs/1103.5733}{{\tt 1103.5733}}].

\bibitem{Berman:2011cg}
D.~S. Berman, H.~Godazgar, M.~Godazgar and M.~J. Perry,  {\em {The local
  symmetries of M-theory and their formulation in generalised geometry}}, JHEP
  {\bf 1201}, 012 (2012)
[\href{http://www.arXiv.org/abs/1110.3930}{{\tt 1110.3930}}].

\bibitem{Berman:2011jh}
D.~S. Berman, H.~Godazgar, M.~J. Perry and P.~West,  {\em {Duality invariant
  actions and generalised geometry}}, JHEP {\bf 1202}, 108 (2012)
[\href{http://www.arXiv.org/abs/1111.0459}{{\tt 1111.0459}}].

\bibitem{Coimbra:2011ky}
A.~Coimbra, C.~Strickland-Constable and D.~Waldram,  {\em {$E_{d(d)} \times
  \mathbb{R}^+$ generalised geometry, connections and M-theory}}, JHEP {\bf
  1402}, 054 (2014)
[\href{http://www.arXiv.org/abs/1112.3989}{{\tt 1112.3989}}].

\bibitem{Coimbra:2012af}
A.~Coimbra, C.~Strickland-Constable and D.~Waldram,  {\em {Supergravity as
  generalised geometry II: $E_{d(d)} \times \mathbb{R}^+$ and M-theory}}, JHEP
  {\bf 1403}, 019 (2014)
[\href{http://www.arXiv.org/abs/1212.1586}{{\tt 1212.1586}}].

\bibitem{Park:2014una}
J.-H. Park and Y.~Suh,  {\em U-gravity: ${\rm SL}(N)$}, JHEP {\bf 1406}, 102 (2014)
[\href{http://www.arXiv.org/abs/1402.5027}{{\tt 1402.5027}}].

\bibitem{Park:2013gaj}
J.-H. Park and Y.~Suh,  {\em {U-geometry: ${\rm SL}(5)$}}, JHEP {\bf 1304}, 147 (2013)
[\href{http://www.arXiv.org/abs/1302.1652}{{\tt 1302.1652}}].

\bibitem{Cederwall:2013naa}
M.~Cederwall, J.~Edlund and A.~Karlsson,  {\em {Exceptional geometry and tensor
  fields}}, JHEP {\bf 1307}, 028 (2013)
[\href{http://www.arXiv.org/abs/1302.6736}{{\tt 1302.6736}}].

\bibitem{Cederwall:2013oaa}
M.~Cederwall,  {\em {Non-gravitational exceptional supermultiplets}}, JHEP {\bf
  1307}, 025 (2013)
[\href{http://www.arXiv.org/abs/1302.6737}{{\tt 1302.6737}}].

\bibitem{Aldazabal:2013mya}
G.~Aldazabal, M.~Gra\~na, D.~Marqu\'es and J.~A.~Rosabal,  {\em {Extended geometry and
  gauged maximal supergravity}}, JHEP {\bf 1306}, 046 (2013)
[\href{http://www.arXiv.org/abs/1302.5419}{{\tt 1302.5419}}].

\bibitem{Aldazabal:2013via}
G.~Aldazabal, M.~Gra\~na, D.~Marqu\'es and J.~A. Rosabal,  {\em {The gauge
  structure of exceptional field theories and the tensor hierarchy}}, JHEP {\bf
  1404}, 049 (2014)
[\href{http://www.arXiv.org/abs/1312.4549}{{\tt 1312.4549}}].

\bibitem{Hohm:2013vpa}
O.~Hohm and H.~Samtleben,  {\em {Exceptional field theory I: $E_{6(6)}$
  covariant form of M-Theory and type IIB}}, Phys.\ Rev.\ {\bf D89}, 066016 (2014)
[\href{http://www.arXiv.org/abs/1312.0614}{{\tt 1312.0614}}].

\bibitem{Hohm:2013uia}
O.~Hohm and H.~Samtleben,  {\em {Exceptional field theory II: $E_{7(7)}$}},
  Phys.\ Rev.\ {\bf D89}, 066017 (2014)
[\href{http://www.arXiv.org/abs/1312.4542}{{\tt 1312.4542}}].

\bibitem{Hohm:2014fxa}
O.~Hohm and H.~Samtleben,  {\em {Exceptional field theory III: $E_{8(8)}$}},
  Phys.\ Rev.\ {\bf D90}, 066002 (2014)
[\href{http://www.arXiv.org/abs/1406.3348}{{\tt 1406.3348}}].

\bibitem{Hohm:2014qga}
O.~Hohm and H.~Samtleben,  {\em {Consistent Kaluza-Klein truncations via
  exceptional field theory}}, JHEP {\bf 01}, 131 (2015)
[\href{http://www.arXiv.org/abs/1410.8145}{{\tt 1410.8145}}].

\bibitem{Rosabal:2014rga}
J.~A. Rosabal,  {\em {On the exceptional generalised Lie derivative for
  $d\geq7$}},
[\href{http://www.arXiv.org/abs/1410.8148}{{\tt 1410.8148}}].

\bibitem{Cederwall:2015ica}
M.~Cederwall and J.~Rosabal,  {\em {$E_8$ geometry}},
JHEP {\bf 1507}, 007 (2015)
[\href{http://www.arXiv.org/abs/1504.04843}{{\tt 1504.04843}}].

\bibitem{Cremmer:1998px}
E.~Cremmer, B.~Julia, H.~L\"u and C.~Pope,  {\em {Dualization of dualities II:
  Twisted self-duality of doubled fields and superdualities}}, Nucl.\ Phys.\ {\bf
  B535}, 242--292 (1998)
[\href{http://www.arXiv.org/abs/hep-th/9806106}{{\tt hep-th/9806106}}].

\bibitem{HenryLabordere:2002dk}
P.~Henry-Labord\`ere, B.~Julia and L.~Paulot,  {\em {Borcherds symmetries in
  M-theory}}, JHEP {\bf 0204}, 049 (2002)
[\href{http://www.arXiv.org/abs/hep-th/0203070}{{\tt hep-th/0203070}}].

\bibitem{HenryLabordere:2002xh}
P.~Henry-Labord\`ere, B.~Julia and L.~Paulot,  {\em {Real Borcherds superalgebras
  and M-theory}}, JHEP {\bf 0304}, 060 (2003)
[\href{http://www.arXiv.org/abs/hep-th/0212346}{{\tt hep-th/0212346}}].

\bibitem{Henneaux:2010ys}
M.~Henneaux, B.~L. Julia and J.~Levie,  {\em {$E_{11}$, Borcherds algebras and
  maximal supergravity}}, JHEP {\bf 1204}, 078 (2012)
[\href{http://www.arXiv.org/abs/1007.5241}{{\tt 1007.5241}}].

\bibitem{Greitz:2011da}
J.~Greitz and P.~Howe,  {\em {Maximal supergravity in $D=10$: Forms, Borcherds
  algebras and superspace cohomology}}, JHEP {\bf 1108}, 146 (2011)
[\href{http://www.arXiv.org/abs/1103.5053}{{\tt 1103.5053}}].

\bibitem{Greitz:2012vp}
J.~Greitz and P.~Howe,  {\em {Half-maximal supergravity in three dimensions:
  supergeometry, differential forms and algebraic structure}}, JHEP {\bf 1206},
  177 (2012)
[\href{http://www.arXiv.org/abs/1203.5585}{{\tt 1203.5585}}].

\bibitem{Greitz:2013pua}
J.~Greitz, P.~Howe and J.~Palmkvist,  {\em {The tensor hierarchy simplified}},
Class.\ Quant.\ Grav.\  {\bf 31}, 087001 (2014)
[\href{http://www.arXiv.org/abs/1308.4972}{{\tt 1308.4972}}].

\bibitem{Howe:2015hpa}
P.~Howe and J.~Palmkvist,  {\em {Forms and algebras in (half-)maximal
  supergravity theories}},
JHEP {\bf 1505}, 032 (2015)
[\href{http://www.arXiv.org/abs/1503.00015}{{\tt 1503.00015}}].

\bibitem{Palmkvist:2013vya}
J.~Palmkvist,  {\em {The tensor hierarchy algebra}},
J.\ Math.\ Phys.\  {\bf 55}, 011701 (2014)
[\href{http://www.arXiv.org/abs/1305.0018}{{\tt 1305.0018}}].

\bibitem{Riccioni:2007au}
F.~Riccioni and P.~C. West,  {\em {The $E_{11}$ origin of all maximal
  supergravities}}, JHEP {\bf 0707}, 063 (2007)
[\href{http://www.arXiv.org/abs/0705.0752}{{\tt 0705.0752}}].

\bibitem{Bergshoeff:2007qi}
E.~A. Bergshoeff, I.~De~Baetselier and T.~A. Nutma,  {\em {$E_{11}$ and the
  embedding tensor}}, JHEP {\bf 0709}, 047 (2007)
[\href{http://www.arXiv.org/abs/0705.1304}{{\tt 0705.1304}}].

\bibitem{Bergshoeff:2007vb}
E.~A. Bergshoeff, J.~Gomis, T.~A. Nutma and D.~Roest,  {\em {Kac-Moody spectrum
  of (half-)maximal supergravities}}, JHEP {\bf 0802}, 069 (2008)
[\href{http://www.arXiv.org/abs/0711.2035}{{\tt 0711.2035}}].

\bibitem{Riccioni:2007ni}
F.~Riccioni and P.~C. West,  {\em {$E_{11}$-extended spacetime and gauged
  supergravities}}, JHEP {\bf 0802}, 039 (2008)
[\href{http://www.arXiv.org/abs/0712.1795}{{\tt 0712.1795}}].

\bibitem{West:2001as}
P.~C. West,  {\em {$E_{11}$ and M-theory}}, Class.\ Quant.\ Grav. {\bf 18},
  4443--4460 (2001)
[\href{http://www.arXiv.org/abs/hep-th/0104081}{{\tt hep-th/0104081}}].

\bibitem{Palmkvist:2011vz}
J.~Palmkvist,  {\em {Tensor hierarchies, Borcherds algebras and $E_{11}$}}, JHEP
  {\bf 1202}, 066 (2012)
[\href{http://www.arXiv.org/abs/1110.4892}{{\tt 1110.4892}}].

\bibitem{Palmkvist:2012nc}
J.~Palmkvist,  {\em {Borcherds and Kac-Moody extensions of simple
  finite-dimensional Lie algebras}}, JHEP {\bf 1206}, 003 (2012)
[\href{http://www.arXiv.org/abs/1203.5107}{{\tt 1203.5107}}].

\bibitem{Berman:2012vc}
D.~S. Berman, M.~Cederwall, A.~Kleinschmidt and D.~C. Thompson,  {\em {The
  gauge structure of generalised diffeomorphisms}}, JHEP {\bf 1301}, 064 (2013)
[\href{http://www.arXiv.org/abs/1208.5884}{{\tt 1208.5884}}].

\bibitem{Cederwall:2015oua}
M.~Cederwall and J.~Palmkvist,  {\em {Superalgebras, constraints and partition
  functions}}, JHEP {\bf 1508}, 036 (2015)
[\href{http://www.arXiv.org/abs/1503.06215}{{\tt 1503.06215}}].

\bibitem{deWit:2005hv}
B.~de~Wit and H.~Samtleben,  {\em {Gauged maximal supergravities and
  hierarchies of non-abelian vector-tensor systems}}, Fortsch.\ Phys.\ {\bf 53},
  442--449 (2005)
[\href{http://www.arXiv.org/abs/hep-th/0501243}{{\tt hep-th/0501243}}].

\bibitem{deWit:2008ta}
B.~de~Wit, H.~Nicolai and H.~Samtleben,  {\em {Gauged supergravities, tensor
  hierarchies, and M-theory}}, JHEP {\bf 0802}, 044 (2008)
[\href{http://www.arXiv.org/abs/0801.1294}{{\tt 0801.1294}}].

\bibitem{Kac77B}
V.~G. Kac,  {\em {L}ie superalgebras}, Adv. Math. {\bf 26}, 8--96 (1977).

\bibitem{Kac}
V.~G. Kac, {\em Infinite dimensional {L}ie algebras}.
\newblock 3rd edition, Cambridge University Press, 1990.

\bibitem{Wakimoto}
M.~Wakimoto, {\em Infinite-dimensional {L}ie algebras}.
\newblock American Mathematical Society, 2001.

\bibitem{Ray}
U.~Ray, {\em Automorphic forms and {L}ie superalgebras}.
\newblock Springer, Dordrecht, 2006.

\bibitem{Kleinschmidt:2013em}
A.~Kleinschmidt and J.~Palmkvist,  {\em {Oxidizing Borcherds symmetries}}, JHEP
  {\bf 1303}, 044 (2013)
[\href{http://www.arXiv.org/abs/1301.1346}{{\tt 1301.1346}}].

\bibitem{Kantor3.5}
I.~L. Kantor,  {\em Some generalizations of {J}ordan algebras}, Trudy Sem.\
  Vect.\ Tens.\ Anal.\ {\bf 16}, 407--499 (1972).

\bibitem{West:2011mm}
P.~West,  {\em {Generalised geometry, eleven dimensions and $E_{11}$}}, JHEP {\bf
  02}, 018 (2012)
[\href{http://www.arXiv.org/abs/1111.1642}{{\tt 1111.1642}}].

\bibitem{West:2012qz}
P.~West,  {\em {$E_{11}$, generalised space-time and equations of motion in four
  dimensions}}, JHEP {\bf 12}, 068 (2012)
[\href{http://www.arXiv.org/abs/1206.7045}{{\tt 1206.7045}}].

\bibitem{West:2014eza}
P.~West,  {\em {Generalised space-time and gauge transformations}}, JHEP {\bf
  08}, 050 (2014)
[\href{http://www.arXiv.org/abs/1403.6395}{{\tt 1403.6395}}].

\end{thebibliography}

\providecommand{\href}[2]{#2}\begingroup\raggedright\endgroup

\end{document}